\documentclass{article}
\usepackage[utf8]{inputenc}
\emergencystretch=\maxdimen
\usepackage[utf8]{inputenc}
\usepackage[hidelinks]{hyperref}
\usepackage{xcolor}
\usepackage{soul}
\usepackage{graphicx}
\usepackage{bm}
\usepackage{float}
\usepackage{mdframed}
\usepackage{algorithm}
\usepackage{algorithmicx}
\usepackage{algpseudocode}
\usepackage[english]{babel}

\newtheorem{objective}{Objective}
\newtheorem{definition}{Definition}

\usepackage{amsmath}
\usepackage{authblk}
\usepackage{amssymb}
\emergencystretch=\maxdimen
\usepackage{draftwatermark}
\SetWatermarkText{DRAFT}
\SetWatermarkLightness{0.96}
\SetWatermarkScale{3}

\newcommand\blfootnote[1]
{
  \begingroup
  \renewcommand\thefootnote{}\footnote{#1}
  \addtocounter{footnote}{-1}
  \endgroup
}
\usepackage{longtable}
\title{IntraLayer: A Platform of Digital Finance Platforms}
\author{Arman Abgaryan$^{I, II}$, Utkarsh Sharma$^{I, III}$}

\date{June 2024}

\begin{document}

\maketitle
\blfootnote{I. Supra}
\blfootnote{II. Shanghai Jiao Tong University}
\blfootnote{III. University of Oxford}
\blfootnote{The authors thank Joshua Tobkin for valuable discussions and feedback on this paper.}

\begin{abstract}    
    IntraLayer presents an innovative framework that enables comprehensive interconnectivity in digital finance. The proposed framework comprises a core underlying infrastructure and an overarching strategy to create a pioneering \lq\lq platform of platforms\rq\rq, serving as an algorithmic fiduciary. By design, this infrastructure optimises transactional efficiency for a broad spectrum of agents, thereby facilitating the sustainable creation of intrinsic economic value. Complementing the infrastructure, our forthcoming work will present an overarching adaptive fiscal policy to optimise IntraLayer's resources, striking a balance between sustaining the network and enhancing the proposal herein.
\end{abstract}

\section{Introduction}
    Blockchain is a shared, immutable ledger that records transactions and tracks asset ownership across a distributed network. It underpins the burgeoning field of peer-to-peer financial systems encapsulated within Decentralised Finance (DeFi) \cite{werner2021sok}, which is based on the permissionless and decentralised nature of blockchain technology, where decentralised applications (dApps) reside. In essence, dApps and blockchain infrastructure enable unrestricted financial accessibility and trustless execution of capital allocation. However, the composability and financial interconnectivity of blockchain-based infrastructures are inherently limited by their ecosystems — a design trait that inadvertently limits a blockchain network's true potential, especially in an expansive and interconnected financial market.\\
    \\
    The infrastructure proposed in this work will be instrumental in digital finance, as resolving the constraints in the disjointed blockchain world not only creates financial value by making capital more efficient but also supports a revolutionary shift from reliance on centralised systems, which inhibit market efficiency, towards the development of decentralised markets atop distributed, global, and permissionless databases. This requires communication channels to seamlessly facilitate synchronisation and value exchange between ecosystems, by resolving fundamental discontinuities, which is envisioned to liberate constrained transactional volumes.\\
    \\
    Adopting an interoperability-centric approach is crucial for realising this vision, as optimising operational efficiency helps enhance the utility of blockchain technologies. The proposed infrastructure reduces the initial costs and complexities associated with establishing inter-platform linkages — spanning diverse blockchain platforms and bridging off-chain/on-chain system dynamics, and consolidates liquidity, reducing the need for participants to create bilateral connections between multiple systems. It revitalises competition by alleviating switching costs, circumventing lock-in effects, and fostering more cohesive, integrated solutions. Consequently, this enhanced interoperability not only improves efficiency and reduces costs, but also enriches the end-user experience, thereby broadening access and inclusivity. In general, this advancement in interoperability is poised to catalyse international economic and financial integration, marking a paradigm shift in furthering interconnectivity between global financial infrastructures.\\
    \\
    The remainder of this paper\footnote{Some notations and optimisation statements in this work are introduced solely to establish a framework for a future, extended version that will include an on-chain solver for core optimisation problems.} is organised as follows: Section \ref{sec:litreview} provides a summary of the body of works that are relevant to the infrastructure proposed herein; Section \ref{sec:obj} provides a comprehensive list of objectives that govern our system architecture; Section \ref{sec:sys} describes our system architecture; Section \ref{sec:utility} explains the utility of the proposed system architecture; Section \ref{sec:business} outlines our business model, encapsulating various strategies to enable system's adoption and fiscal sustainability addressing the network's strategic initiatives and economic objectives; Section \ref{sec:conclusions} provides a concluding summary and motivates future work; after which the appendix provides a handy list of notations, and our vision for how the infrastructure presented herein can serve as a gateway to connect decentralised and centralised finance, promoting efficiency and synergy across distinct sectors of the digital and traditional financial community.

\section{Literature Review}\label{sec:litreview}
    This work builds upon the literature in digital finance, focusing on interoperability, gateway design, network business models, inter-blockchain data connectivity, and the nascent field of digital assets. Among numerous pertinent studies, this paper concentrates on a subset that can be categorised as follows:   
    \begin{itemize}
        \item \textbf{On-Chain Financial Systems}: The IntraLayer infrastructure seeks to propel the on-chain financial system's evolution, a swiftly growing segment of the digital finance ecosystem. On-chain finance includes a variety of financial products such as cryptocurrencies lending markets \cite{bartoletti2020, Xu2022}, decentralised exchanges \cite{Xu2023, mohan2022}, and even though not necessarily permissionless, it can also include central bank digital currencies (CBDCs) \cite{Fernandez2021, allen2020}.
        
        \item \textbf{Gateways}: Advancing the discourse on gateways, the IntraLayer infrastructure enhances interactions among complex systems. Gateways are instrumental in tackling system fragmentation \cite{david1988, david1987}, where even imperfect gateways aid incompatibility issues \cite{farrell1992}. They facilitate product compatibility and interoperability, expanding user options \cite{matutes1988}, and enabling new market entrants \cite{kades2020}. Recent scholarly interest in cross-chain interoperability focuses on essential attributes of Distributed Ledger Technology (DLT) gateways \cite{hardjono2020}, and cross-chain communication protocols \cite{mohanty2020, belchior2021, zamyatin2019}. The IntraLayer system parallels traditional Enterprise Service Bus (ESB) \cite{menge2007} features, functioning as a digital finance ESB, providing integration services and ensuring interoperability among heterogeneous systems. For data routing, it mirrors message brokers \cite{longo2022}, embodying a decentralised message broker architecture.
        
        \item \textbf{Decentralised Network Business Models}: The IntraLayer infrastructure presents an opportunity to execute a new business model within the decentralised network sphere, which differs from the business models adopted existing networks specialising in various domains such as: smart contracts \cite{ethereum, solana, cardano}, privacy \cite{monero, zcash, secret}, payments \cite{bitcoin, litecoin, stellar}, identity \cite{ontology, sovrin}, interoperability \cite{wanchain, polkadot, cosmos}, and storage \cite{filecoin, arweave, storj}. We introduce a decentralised network uniquely designed for financial interconnectivity and transactional automation, bridging on-chain and off-chain systems, supported by a business model that fosters network sustainability and evolution.
        
        \item \textbf{Interoperability}: As an increasingly pivotal theme in digital finance, interoperability studies include market analyses \cite{arabehety2016}, CBDC exchange models \cite{Jung2021, Auer2020}, etc. The IntraLayer infrastructure is akin to traditional clearinghouses \cite{koeppl2012}, reducing costs and complexities in clearing and settlement, thereby promoting interoperability.
    \end{itemize}

\section{Objectives}\label{sec:obj}
    We model the digital finance ecosystem as a directed graph $\mathbf{G} = (\mathbf{V}, \mathbf{E})$, where the vertices $\mathbf{V} \subseteq \mathbb{N}$ represent a finite set of individual agents within the ecosystem, including dApps, end-users, and other economic agents, whether on or off a blockchain network. The edges $E \subseteq V \cdot (V-1)$ encapsulate the transactional paths among these agents, such that each $v_i \in V$ corresponds to a specific agent. Therefore, the set of edges is defined as $\mathbf{E} = \{(v_i, v_j) \mid v_i, v_j \in \mathbf{V}, i \neq j \} $, where each directed edge $(v_i, v_j) \in E $ represents a transactional path from agent $v_i$ to agent $v_j$. Note, that each edge $(v_i, v_j)$ may also be associated with a weight $w^{ij}_e \in \mathbb{R}$ representing the value of transactions conducted using a path, thus forming a weighted graph $\mathbf{G} = (\mathbf{V}, \mathbf{E}, \mathbf{W})$, with $W = \{ w^{ij}_e \mid (v_i, v_j) \in E \text{ and } w^{ij}_e \in \mathbb{R} \}$ as the set of interaction weights.

    \begin{definition}[Transactional Path]
        A transactional path is a connection between two agents, say - $a$ and $b$, on either single-platform or cross-platform (i.e., $P_i$ and $P_j$ $\forall i = j$, or $P_i$ and $P_j$ $\forall i \neq j$, respectively), which enables financial transactions and is characterised by a specific epoch's (a fixed time interval - $e$) economic output ($O^{a,b}_e \in \mathbb{R}$) and cost ($C^{a,b}_e \in \mathbb{R}_{\geq 0}$) - which is influenced by factors such as data connectivity, value conversion, value connectivity, clearing and setup complexity, distinguished by subscript, as follows:
        \begin{equation*}
            C^{a,b}_e = f_1(C^{a,b}_{k_e}) \forall k,
        \end{equation*}
        \noindent
        where, $k \in \{\text{DC=DataConnectivity}, \text{VT=ValueConnectivity}, \text{VC=ValueConversion},$\\
        $\text{CS=Clearing}, \text{SC=SetupComplexity}\}$, $f_1: \mathbb{R}^5 \to \mathbb{R}_{\geq 0}$ is a function mapping a vector of epoch-specific cost for agents $a$ and $b$ ($C^{a,b}_{k_e} \in \mathbb{R}_{\geq 0}$) to the overall cost of setting up transactional paths.
    \end{definition}

    \noindent
    Note, that although IntraLayer can also be used for transactional paths on a single-platform, it is primarily designed to enable the formation of cross-platform transactional paths.\\
    \\
    In essence, IntraLayer's approach to achieving its multi-faceted objective of forming cross-platform transactional paths can be envisioned as an optimisation problem, as follows:

    \begin{align*}
        &\text{Maximise}_{\bm{\mathcal{P}}, \bm{I}_u, \bm{F}_u, \bm{\mathcal{K}}_u, \bm{\Theta}_u, \mathbf{B}_{\Omega_u}}: \sum^{\mathcal{E}}_{u=1} \mathbb{E}\left[ \sum_{\substack{i=1 \\ i \neq j}}^{N_a} \sum^{N_a}_{j=1} \left( \mathbf{O^{i,j}_u}(\bm{\mathcal{P}}, \bm{\mathcal{K}}_u, \bm{\Theta}_u) - \mathbf{C^{i,j}_u}(\bm{\mathcal{P}}, \bm{I}_u, \bm{F}_u, \bm{\mathcal{K}}_u, \mathbf{B}_{{\Omega_u}}) - \beta_u \cdot (R_u - \Gamma_u) \right) \right],\\
        &\text{Subject to:} 
    \end{align*}
    \begin{equation}
    \begin{aligned}
        & \mathbf{D}_{u + \Delta u} = g_{\mathbf{D}}(\mathbf{D}_{0:u}, \kappa_{0:u}, \dots) \quad \forall \, \mathbf{D} \in \{ \mathbf{I}, \mathbf{F}, \bm{\mathcal{K}}, \bm{\Theta}, \mathbf{B}_{\Omega} \},
    \end{aligned}
    \end{equation}

    \noindent
    where $\mathcal{E} \in \mathbb{N}$ is a system parameter that describes the length of the period during which the IntraLayer network aims to achieve its objectives; $\bm{\mathcal{P}}$ represents a set of design sets, each of which are \textit{presently} of unspecified dimensions; $\bm{I} \in \mathbb{R}^{m \cdot \mathcal{E} \cdot N_A}_{\geq 0}$ is the tensor of system's incentives (expenditures) for $m$ services across epochs and asset pools; $\mathbf{B}_{\Omega} \in \mathbb{R}_{\geq 0}^{3\mathcal{E}}$ is a matrix which represents the budget allocated to compensate the service providers for operating the technical infrastructure; $\bm{F} \in \mathbb{R}^{m \cdot \mathcal{E}}_{\geq 0}$ is the matrix of fees (revenue) levied on customers across $m$ services and epochs; $N_a \in \mathbb{N}$ is the number of agents involved in the system; $\mathbf{O^{i,j}_u}: (\mathbb{R}^{{d_x}}, \mathbb{R}^{2\cdot N_A},\mathbb{R}^{{d_x}})\rightarrow \mathbb{R}$ is a function of economic output derived by agents $i$ and $j$ from the transactional path connecting them in $u$-th epoch and $d_x$ being a placeholder for any notation of unknown dimension; $R_u \in \mathbb{R}_{\geq 0}$ signifies the total resources of the system, which from an accounting perspective represents the equity (assets minus liabilities) available in the system's balance sheet for $u$-th epoch; $\Gamma_u \in \mathbb{R}_{\geq 0}$ is a system parameter signifying the amount of system-resources that are required in each epoch; $ \mathbf{C^{i,j}_u}: (\mathbb{R}^{{d_x}},\mathbb{R}^{m \cdot N_A},  \mathbb{R}^{m},  \mathbb{R}^{2 \cdot N_A}, \mathbb{R}^3_{\geq 0} ) \rightarrow \mathbb{R}_{\geq 0}$ represents the cost function for agents' ($i$ and $j$) interaction at the $u$-th epoch; $\bm{\mathcal{K}} \in \mathbb{R}^{2\cdot N_A \cdot \mathcal{E}}$ is the matrix of the portfolio of ($N_A$ many) assets controlled by the system across different epochs, allocated for different economic purposes; $\bm{\Theta} \in \mathbb{R}^{{d_x}}$ is a vector of sets representing the agent acquisition costs allocated for acquiring different agents into the system across various epochs; $\beta_u$ is some weight; $\mathbf{\kappa}_u \in \mathbb{R}^{{d_x}}$ is a set of vectors representing state variables at $u$-th epoch; $g_{\mathbf{D}}: \mathbb{R}^{{d_x}} \rightarrow \mathbb{R}^{{d_x}}$ is the transition function evolving the decision variables from epoch $u$ to epoch $u + \Delta u$ such that $\Delta u$ represents the discrete time interval. Note that when a matrix intended to represent values for multiple epochs is presented with an epoch-specific subscript, it refers to the elements of that matrix corresponding to the referenced epoch.

\subsection{Cost}
    Having outlined IntraLayer's core summary objective in the preceding section, we now proceed to describing associated sub-objectives in the rest of this section.
    
    \begin{objective}[Data Integration]
        IntraLayer seeks to be the robust infrastructure for seamless data exchange between agents, paving the way for robust and widespanning financial interconnectivity.\\
        \\
        This is in response to fragmentation in current digital finance landscape, which is characterised by isolated databases and a lack of reliable infrastructure for data exchange between disparate systems, impeding development of effective transactional paths, which are crucial for enhancing financial interconnectivity and promoting economic stability and growth in digital finance. This is further exasperated by the lack of trustless access to the state information of other agents, whether these are based on other blockchain platforms or off-chain systems, limiting the potential of blockchain technology to serve as the foundation of a global financial system.\\
        \\
        For a set of platforms $\mathbf{P} = \{P_1, P_2, \ldots, P_n\}$, and agents $\mathbf{a} = \{a_{i1}, a_{i2}, \ldots, a_{im}\}$ operating on platform $P_i$, which may be smart contracts (characterised by state data), off-chain systems, or users, the cost of data connectivity between agents $a_{ik}$ and $a_{jk}$ is denoted as $C_{{DC}_e}^{{a_{ik}, a_{jk}}} \in \mathbb{R}_{\geq 0}$. This cost, which facilitates transactional relationships, depends on the risk of data quality compromise and any system-imposed fees, and as such, the cost increases if data is stored in disjoint systems. However, solving this requires satisfactorily addressing the following two key factors:
        
        \begin{itemize}
            \item \textbf{Accuracy}: If $\mathcal{D}^{ik}_{{jk}_t} \in \mathbb{R}^{{d_x}}$ represents agent $a_{ik}$'s data available to agent $a_{jk}$, and $D^{{\star}^{ik}}_{{jk}_t}$ is the data necessary for $a_{jk}$ to efficiently transact with $a_{ik}$, communication costs rises if $D^{ik}_{{jk}_t}$ is a proper subset of (not all necessary data is available) $D^{{\star}^{ik}}_{{jk}_t}$ , or if $\mathcal{D}_{{jk}_t}^{ik} \nsubseteq D^{{\star}^{ik}}_{{jk}_t}$ (communicated data contains inaccuracies). The discrepancy between the received data ($\mathcal{D}^{ik}_{{jk}_t}$) and the relevant data ($D^{{\star}^{ik}}_{{jk}_t}$) is denoted by $\Psi^{ik}_{{jk}_t} \in \mathbb{R}$. 
            
            \item \textbf{On-time Communication}: The communication lag, denoted by $d_{ik, jk} \in \mathbb{R}_{+}$, represents the average delay between the time data is needed by agent $a_{jk}$ ($t_{ik,jk}^\star$) and when it is received ($t_{ik,jk}$) during an epoch. Mathematically:

            \begin{equation}
                d_{{ik, jk}_e} = \frac{1}{n} \sum_{h=1}^{n} (t_{{ik,jk}_e}^h - t_{{ik,jk}_e}^{\star h}) > 0,
            \end{equation}
            
            \noindent
            where $n$ denotes the count of communication events between agents. This delay incurs additional costs, including the economic cost of value loss due to the data communication lag.
        \end{itemize}

        \noindent
        INtraLayer's strength stems from the 
        IntraLayer uses the following three tools to address the need for both accuracy and timely communication:
        
        \begin{itemize}
            \item \textbf{Parametric Design}: IntraLayer's design choices relating to the technical infrastructure, which enables data communication, are encapsulated in a vector of design parameters $(\bm{\mathcal{P}}_{DC} \in \mathbb{R}^{\eta} \text{where } \eta \in \mathbb{N})$.            
            \item \textbf{Economic Security Guarantees}: IntraLayer seeks to attract guarantors, who are special agents using their capital as collateral to guarantee the quality of data communication between agents, and deter malicious activities. To attract multi-asset collateral from agents to secure the system, economic incentives are disbursed denoted by matrix $\mathbf{I}_{\mathbf{S}^{DC}} \in \mathbb{R}^{N_A \cdot \mathcal{E}}_{\geq 0}$. This matrix represents the allocation of incentives to attract different assets across various epochs.            
            \item \textbf{Budget Allocation}: The budget allocated by IntraLayer to compensate service providers operating the value connectivity infrastructure, is captured in the budget allocation vector - $\mathbf{B_{{\Omega}^{DC}}} \in \mathbb{R}_{\geq 0}^{\mathcal{E}}$. This allocation of budget creates economic incentives for service providers to aspire to provide a high quality of service.
        \end{itemize}
 
        \noindent      
        Combined together, and along with an epoch-specific data facilitation fee charged from users ($\mathbf{F}_{DC} \in \mathbb{R}^{\mathcal{E}}_{\geq 0}$), we can express combined costs as follows:
                
        \begin{equation}
            C_{{DC}_e}=\mathbf{f_2}( \bm{\Psi}_e(\bm{\mathcal{P}}_{DC}, \mathbf{I}_{{S^{DC}_e}}, {B}_{{{\Omega}^{VC}_e}})= \sum^{N_a}_{i=1} \sum^{N_a}_{j=1 \neq i} \Psi^i_{j_e}, \mathbf{d}_e (\bm{\mathcal{P}_{DC}}, \mathbf{I}_{S^{DC}_e}, {B}_{{\Omega}^{VC}_e}) = \sum^{N_a}_{i=1} \sum^{N_a}_{j=1 \neq i} d_{{i,j}_e}, F_{{DC}_e}),
        \end{equation}  

        \noindent
        where $\bm{\Psi}_e: (\mathbb{R}^{{d_x}},  \mathbb{R}_{\geq 0}^{N_A \cdot \mathcal{E}}, \mathbb{R}_{\geq 0}) \to \mathbb{R}$ is a function that maps the input values to the total data discrepancy observed in an epoch; $\mathbf{d}_e: (\mathbb{R}^{{d_x}}, \mathbb{R}_{\geq 0}^{N_A \cdot \mathcal{E}}, \mathbb{R}_{\geq 0}) \to \mathbb{R}_{+}$ is the communication lag function that maps the input values to the total communication lag observed in epoch $e$; $F_{{DC}_e}$ is the fee charged by the infrastructure for data connectivity; the function $f_2: (\mathbb{R}_{\geq 0}, \mathbb{R}_{+}, \mathbb{R}_{\geq 0}) \to \mathbb{R}$ outputs the cost of data connectivity; and $\mathbf{I}_{S^{DC}_e} \in \mathbb{R}^{N_A}_{\geq 0}$ represents a vector (sub-matrix) showing the allocation of incentives (expenditures for the network) to attract different assets within a particular epoch.\\
        \\
        In essence, IntraLayer's objectives associated with a multi-period minimisation of the cost of data communication can be expressed as follows:

        \begin{align*}
            &\text{Minimise}_{\bm{\mathcal{P}_{DC}}, \mathbf{I_{S^{DC}_u}}}: \sum^{\mathcal{E}}_{u=1} \mathbb{E} \left[ C_{{DC}_u}(\bm{\mathcal{P}_{{DC}}}, \mathbf{I_{S^{DC}_u}}, \mathbf{B}_{{\Omega_u^{DC}}}, \mathbf{F_{{DC_ u}}}) \right] \\
            &\text{Subject to:}
        \end{align*}

        \begin{equation}
        \begin{aligned}
            &\sum^{N_A}_{n=1} I^n_{S^{DC}_u}  =  B_{S^{DC}_u}\\
            &\mathbf{D}_{u + \Delta u} = g_{\mathbf{D}}(\mathbf{D}_{0:u}, \mathbf{\kappa}_{0:u}, \dots) \quad \forall \, \mathbf{D} \in \{\mathbf{I}_{S^{DC}}\},
        \end{aligned}
        \end{equation}

        \noindent
        where $N_A$ is the number of assets in the system; $B_{S^{DC}_u} \in \mathbb{R}_{\geq 0}$ is the epoch-specific budget of incentives to attract capital that is applied to secure the data connectivity enabled by the infrastructure; and $I^n_{S^{DC}_u}$ is the incentive amount distributed to attract
        $n$-th asset. 
    \end{objective}

    \begin{objective}[Value Integration]
        IntraLayer seeks to overcome the challenge of fragmented pods of value, leading to agents incurring a higher cost of value transfer, which not only includes the operational cost of value interconnectivity, but also heightened security costs stemming from the risk of value being lost during transfer.\\
        \\
        This cost can be calculated as:    
        \begin{equation*} 
            C_{{VT}_e}^{a_{ik}, a_{jk}} = NF_{{VT}_e}^{a_{ik}, a_{jk}} + \mathbb{E}[C_{{Security}_{e}}^{a_{ik}, a_{jk}}],
        \end{equation*}
        
        \noindent
        where $NF_{{VT}_e}^{a_{ik}, a_{jk}} \in \mathbb{R}_{\geq 0}$ represents the fee paid by agents to transfer value from agent $a_{ik}$ to agent $a_{jk}$; and $\mathbb{E}[C_{\text{Security}_e}^{a_{ik}, a_{jk}}]$ is expected cost of realising the security risk from transferring value, both expressed in nominal terms and specific to a particular epoch. This expected security cost is a function of the infrastructure design, encapsulated in the vector of system parameters $\bm{\mathcal{P}}_{VT} \in \mathbb{R}^\eta \text{where } \eta \in \mathbb{N}^{+}$, the incentives distributed to attract assets from guarantors to secure value connectivity ($\mathbf{I}_{S^{VT}}$), and the vector of the budget allocated for different epochs to compensate the service providers who facilitate value connectivity ($\mathbf{B}_{{\Omega}^{VT}} \in \mathbb{R}^{\mathcal{E}}_{\geq 0}$).\\
        \\
        If we have assets $\mathbf{A}_i = \{A_{i1}, A_{i2}, \ldots, A_{il}\}$ on blockchain $B_i$, and $\mathbf{a}_i = \{a_{i1}, a_{i2}, \ldots, a_{im}\}$ acting as agents (e.g. dApps, users, etc.) on $B_i$, then the transactional friction of transmitting value across different blockchains due to value fragmentation is represented by the cost efficiency of value connectivity ($CE^{a_{ik},a_{jk}}_{{VT}_e} \in \mathbb{R}_{+}$) between agents ($a_{ik}$ and $a_{jk}$), and calculated as follows:
        
        \begin{equation}
            CE^{a_{ik}, a_{jk}}_{{VT}_e} = \frac{\mathcal{V}^{a_{ik}, a_{jk}}_{{VT}_e}}{C^{a_{ik}, a_{jk}}_{{VT}_e}} \forall C^{a_{ik}, a_{jk}}_{{VT}_e} \neq 0,
        \end{equation}
        
        \noindent
        where $\mathcal{V}^{a_{ik}, a_{jk}}_{{VT}_e} \in \mathbb{R}_{\geq 0}$ is the nominal value transferred from agent $a_{ik}$ to agent $a_{jk}$. A focus on enhancing this measure of cost efficiency enables the network to mitigate value fragmentation and enhances the utility of assets.\\
        \\
        The objective of optimising transaction paths and financial interconnectivity is to maximise the cost efficiency of value transfers for all agents involved in the system:

        \begin{align*}
            &\text{Maximise}_{\bm{\mathcal{P}_{{VT_u}}}, \mathbf{I_{S^{VT}_u}}} \quad  \mathbb{E} \left[ \frac{\sum_{u=1}^{\mathcal{E}} \mathcal{V}_{{VT}_u} \cdot \mathbf{CE}_{{VT}_u}(\bm{\mathcal{P}_{{VT_u}}}, \mathbf{F}_{{VT_u}}, \mathbf{I_{S^{VT}_u}}, \mathbf{B}_{{\Omega}^{VT}_u})}{\sum^{\mathcal{E}}_{u=1} \mathcal{V}_{{VT}_u}} \right]\\
            &\text{Subject to:}
        \end{align*}
        
        \begin{equation}
        \begin{aligned}
            &\sum_{z=1}^{N_A} I^z_{S^{VT}_u} \leq B_{S^{VT}_u},\\
            & \mathbf{D}_{u + \Delta u} = g_{\mathbf{D}}(\mathbf{D}_{0:u}, \mathbf{\kappa}_{0:u}, \dots) \quad \forall \, \mathbf{D} \in \{ \bm{\mathcal{P}}_{VT}, \mathbf{I}_{S^{VT}}\},
        \end{aligned}
        \end{equation}
        
        \noindent
        where $\mathbf{I_{S^{VT}}} \in \mathbb{R}^{N_A \cdot \mathcal{E}}_{\geq 0}$ represents the matrix of economic incentives (expenditures for the system) applied to attract various assets to secure the value connectivity supported by IntraLayer in different epochs; $ \mathbf{F}_{VT} \in \mathbb{R}^\mathcal{E}_{\geq 0}$ is the fee for value transfer; $I^z_{S^{VT}_u}$ is the incentive distributed (in nominal terms) to attract asset $z$ in epoch $u$; $\mathcal{V}_{{VT}_u} \in \mathbb{R}_{+}$ is the total nominal value transferred in $u$-th epoch; $\mathbf{B}_{{\Omega}^{VT}} \in \mathbb{R}_{\geq 0}^{\mathcal{E}}$ represents the budget allocated to cover expenses associated with compensating service providers who operate the value connectivity infrastructure; $B_{S^{VT}_u} \in \mathbb{R}_{\geq 0}$ is the budget of incentives for epoch $u$ to attract capital applied to secure the value connectivity enabled by the infrastructure; $\mathcal{E}$ symbolises the aggregate number of epochs during which the system endeavours to optimise the cost efficiency of value connectivity; and the cost efficiency function that provides the aggregate cost efficiency of value connectivity amongst supported paths - $\mathbf{CE}_{{VT}_u}: (\mathbb{R}^{\eta}, \mathbb{R}^{\mathcal{E}}, \mathbb{R}^{\mathcal{E} \cdot N_A} , \mathbb{R}^{\mathcal{E}}) \to \mathbb{R}_{+} \text{ where } \eta \in \mathbb{N}^{+}$ is:
        \begin{equation}
            \mathbf{CE}_{{VT}_u}(\bm{\mathcal{P}}_{VT}, \mathbf{F}_{VT}, \mathbf{I}_{S^{VT}}, \mathbf{B}_{{\Omega}^{VT}})= \frac{\sum^{N_a}_{g=1 \neq l} \sum^{N_a}_{l=1} \mathcal{V}^{g,l}_{VT_u} \cdot CE^{g,l}_{{VT}_u}}{\sum^{N_a}_{g=1 \neq l} \sum^{N_a}_{l=1}\mathcal{V}^{g,l}_{VT_u}},
        \end{equation}
    \end{objective}
    
    \begin{objective}[Liquidity Integration]
        IntraLayer seeks to integrate fragemented liquidity, which arises due to disparate distribution of capital across liquidity pools, and the resulting inefficiencies stemming from transferring liquidity between them in a cross-chain environment.\\
        \\
        The opportunity presented by realising this objective can be understood in the context of factors which further exacerbate challenges posed by liquidity fragmentation, like -  high capital costs and complexities involved in bootstrapping liquidity for new assets, and the inefficient designs of digital asset marketplaces, making them prone to high slippage and impermanent loss \cite{engel2021}.\\
        \\
        Let $\mathbf{A} = \{A_{11}, A_{12}, \ldots, A_{z'o}\}$ represent the set of assets available in a digital finance ecosystem, where $z'$ represents the number of ecosystems and $o$ the number of assets within $z'$-th ecosystem. We can depict cost-aware system-wide liquidity as a directed acyclic graph $\mathbf{T}(\mathbf{V'}, \mathbf{E'})$, where the vertices $\mathbf{V'}$ represents the assets, and the directed edges - $\mathbf{E'}$, represent the conversion paths between assets. Each edge $E_{il,jl} \in \mathbf{E'}$ from vertex $A_{il}$ to vertex $A_{jl}$ is weighted by the cost efficiency ($CE^{A_{il}, A_{jl}}_{{\text{VC}}_e} \in \mathbb{R}_{+}$) of converting asset $A_{il}$ to $A_{jl}$, which is calculated as follows:
        
        \begin{equation*}
            CE^{A_{il}, A_{jl}}_{{\text{VC}}_e} = \frac{\mathcal{V}^{A_{il}, A_{jl}}_{{\text{VC}}_e}}{C^{A_{il}, A_{jl}}_{{\text{VC}}_e}}  \forall C^{A_{il}, A_{jl}}_{{\text{VC}}_e} \neq 0
        \end{equation*}
        
        \noindent
        where $\mathcal{V}^{A_{il}, A_{jl}}_{{\text{VC}}_e} \in \mathbb{R}_{\geq 0}$ represents the nominal amount of converted volume at a (nominal) cost of $C^{A_{il}, A_{jl}}_{{\text{VC}}_e} \in \mathbb{R}_{+}$.\\
        \\
        As such, IntraLayer seeks to integrate fragmented liquidity in a way that maximises cost efficiency by optimising both the distribution and utilisation of liquidity. This involves using incentives to attract multi-asset capital for liquidity pools and using capital directly controlled by the IntraLayer network, thereby enabling efficient value conversion and reducing slippage. More formally, we can state this as follows:

        \begin{align*}
            &\text{Maximise}_{\bm{\mathcal{P}_{VC}}, \bm{I}_{\mathcal{L}_ u}, \bm{\mathcal{K}}_{{VC_u}}} \quad  \mathbb{E} \left[ \frac{\sum^{\mathcal{E}}_{u=1} \mathcal{V}_{{VC}_u} \cdot \mathbf{CE}_{{\text{VC}}_u}(\bm{I}_{{\mathcal{L}_u}}, \bm{\mathcal{P}_{{VC_u}}}, \bm{\mathcal{K}}_{{VC_u}}, \bm{F}_{{VC_u}})}{\sum^{\mathcal{E}}_{u=1} \mathcal{V}_{{VC}_u}} \right] \forall \mathcal{V}_{{VC}_u} \neq 0\\
            &\text{Subject to:} \\
        \end{align*}
        \begin{equation}
        \begin{aligned}
            &\sum^{N_A}_{o=1}  I_{\mathcal{L}_{o_u}}  \leq B_{\mathcal{L}_u}\\
            &\sum^{N_A}_{o=1}  \mathcal{K}_{{VC}_{o_u}} = \sum^{N_A}_{o=1}  \mathcal{K}_{{VC}_{o_{u-1}}} + B_{{VC}_{\mathcal{K}_u}}\\
            & \mathbf{D}_{u + \Delta u} = g_{\mathbf{D}}(\mathbf{D}_{0:u}, \mathbf{\kappa}_{0:u}, \dots) \quad \forall \, \mathbf{D} \in \{ \bm{I}_{\mathcal{L}}, \bm{\mathcal{K}}_{VC}\},
        \end{aligned}   
        \end{equation}
        
        \noindent
        where $\mathcal{E}$ represents the period during which the system aims to maximise cost efficiency\footnote{The system desires to have a sustainable source of liquidity to achieve its objectives, which requires longer-term availability of multi-asset capital.}; $\bm{I}_{\mathcal{L}} \in \mathbb{R}^{N_A \cdot \mathcal{E}}_{\geq 0}$ is the incentive matrix representing the allocation of incentives to attract liquidity across different liquidity pools and epochs;  $\bm{I}_{\mathcal{L}}$ is a submatrix of $\bm{I}$; $\mathcal{V}_{{VC}_u} \in \mathbb{R}_{+}$ is total nominal amount of value converted in epoch $u$; the  $\bm{\mathcal{P}}_{VC} \in \mathbb{R}^{\eta} \text{where } \eta \in \mathbb{N}^{+}$ represents the vector of design parameters characterising the proposed liquidity aggregation mechanism and asset marketplace as part of the IntraLayer infrastructure, $\bm{\mathcal{P}}_{VC}$ is a subset of $\bm{\mathcal{P}}$; $\bm{F}_{VC} \in \mathbb{R}_{\geq 0}^\mathcal{E}$ is a vector of fees charged by IntraLayer for asset conversion across different epochs, $\bm{F}_{VC}$ is a submatrix of $\bm{F}$; $\bm{\mathcal{K}}_{VC} \in \mathbb{R}^{N_A \cdot \mathcal{E}}_{\geq 0}$ is a matrix of capital (controlled by IntraLayer - Sec. \ref{sec:NOL}) used for value conversion across various liquidity pools and epochs, and $\bm{\mathcal{K}}_{VC} \in \mathbb{R}^{N_A \cdot \mathcal{E}}_{\geq 0}$ is a submatrix of $\bm{\mathcal{K}}$; $N_A$ is the number of assets; $B_{\mathcal{L}_e} \in \mathbb{R}_{\geq 0}$ is the epoch-specific incentive budget for attracting multi-asset liquidity; $\mathcal{K}_{{VC}_{o_u}}$ is the total nominal value of asset $o$ controlled by IntraLayer; $B_{{VC}_{\mathcal{K}_u}} \in \mathbb{R}$ represents the allocated budget to adjust the capital controlled by IntraLayer.\\
        \\
        The function $\mathbf{CE}_{\text{VC}_e}$ is defined as follows:
        
        \begin{equation*}
            \mathbf{CE}_{\text{VC}_e}(\bm{I_\mathcal{L}}, \bm{\mathcal{P}}_{VC}, \bm{\mathcal{K}}_{VC}, \bm{F}_{VC}) = \frac{\sum^{N_A}_{y=1 \neq g} \sum^{N_A}_{g=1} (\mathcal{V}^{A_y, A_g}_{{VC}_e} \cdot CE_{\text{VC}_e}^{A_y, A_g})}{\sum^{N_A}_{y=1\neq g} \sum^{N_A}_{g=1} (\mathcal{V}^{A_y, A_g}_{{VC}_e})},
        \end{equation*}
        \noindent
        where $CE_{\text{VC}_e} \in \mathbb{R}^+$ is value conversion efficiency.
    \end{objective}

    \begin{objective}[Setup Simplicity]
        IntraLayer seeks to simplify the process of setting up transactional paths on the platform, as the complexity of setting up transactional paths in a digital finance ecosystem could adversely impact costs.\\
        \\
        If $C^{a_{ik}, a_{jk}}_{{SC}} \in \mathbb{R}_{\geq 0}$ represents the cost of establishing transactional paths between agents $a_{ik}$ and $a_{jk}$, we can state the cost for $a_{jk}$-th agent to establish bilateral transactional paths with $N_a-1$ other agents involved in the ecosystem, by:
        \begin{equation*}
            C^{a_{jk}, \mathbf{a}}_{\text{SC=SetupComplexity}} = \sum^{N_a-1}_{i=1} C^{a_{jk}, i}_\text{SC}
        \end{equation*}

        \noindent
        Now, if a single gateway entity ($\mathcal{G}$) exists, such that it connects all $N_a$ agents in a hub-and-spoke manner, facilitating the establishment of paths:

        \begin{equation*}
            \exists \mathcal{G} \quad \text{such that} \quad \forall a_i \in \mathbf{a}, \quad a_{jk} \text{ can interact with } a_{i \neq {jk}} \text{ through } \mathcal{G}.
        \end{equation*}

        \noindent
        Such a central gateway entity would decrease the need for multiple smart contracts across different chains and the necessity for bilateral transactional paths between disparate agents, mitigating operational complexities and inefficiencies, as agent $a_{jk}$ would just need to connect to (or be based on) the gateway entity in order to access all other agents. Therefore, if IntraLayer could act as a single, efficient and effective central gateway, it would minimise the setup costs for all agents in the ecosystem, and as such, this objective can be stated as a minimisation problem focused on setup complexity for all agents.

        \begin{equation}
         \begin{aligned}
            &\text{Minimise}_{\bm{\mathcal{P}_{SC}}} \quad \mathbb{E} \left[ \sum^{N_a}_{o=1} C^{o, \mathbf{a}}_{SC_u}(\bm{\mathcal{P}_{SC}}) \right],
        \end{aligned}
        \end{equation}
       
        \noindent
        where $\bm{\mathcal{P}}_{SC} \in \mathbb{R}^{\eta} \forall \eta \in \mathbb{N}^{+}$ and $\bm{\mathcal{P}}_{SC} \subset \bm{\mathcal{P}}$.\\
        \\
        Therefore, IntraLayer seeks to act as a central gateway, enabling seamless integration and interaction across the entire digital finance ecosystem, thereby maximally reducing operational complexity for new agents (e.g. dApps).
    \end{objective}

    \begin{objective}[Unified Clearing]
        IntraLayer seeks to serve as a unified clearing layer, as it consolidates transactions across blockchains and digital finance platforms, reducing costs and complexity for multi-layered settlements.\\
        \\
        Let $\mathbf{TC} = \{\{\mathcal{S}_1, \mathcal{S}_2, \ldots, \mathcal{S}_n\}, \mathbf{EL}\}$ represent a transaction cluster - comprised of a state vector of agents $\mathcal{S}_i \in \mathbb{R}^{{d_x}}$ interacting with the same transactional logic, and associated execution logic ($\mathbf{EL} \in \mathbb{R}^{{d_x}}$). Here, if each agent interacts with every other agent in the network, they would have to receive and process $n \cdot (n-1)$ interactions. We quantify the cost of processing such a transaction cluster as:
        
        \begin{equation*}
            C^{\mathbf{TC}}_{{Clearing}_e} = n \cdot (n-1) \cdot (C^{\mathbf{TC}}_{{DC}_e} + C^{\mathbf{TC}}_{{VC}_e}) + n \cdot C^{\mathbf{TC}}_{{Processing}_e},
        \end{equation*}
        
        \noindent
        $C^{\mathbf{TC}}_{{DC}e}$ and $C^{\mathbf{TC}}_{{VC}_e}$ represent the average costs of data communication and value transfer, respectively, between any two agents within a cluster under a given execution logic; and $C^{\mathbf{TC}}_{{Processing}_e}$ is the average cost for an agent to process settlement.\\
        \\
        By leveraging a single clearing layer as a gateway for all transactions, the system scales from a non-linear interaction model requiring $n \cdot (n-1)$ interactions to a linear model with only $2n-1$, interactions, thereby achieving substantial cost savings. This results in cost savings, which are calculated as follows:
        
        \begin{equation*}
        \begin{aligned}
            \text{CostSavings}^{TC}_e &= n \cdot (n-1) \cdot (C^{\mathbf{TC}}_{{DC}_e} + C^{\mathbf{TC}}_{{VC}_e}) \\
            &\quad + n \cdot C^{\mathbf{TC}}_{{Processing}_e} 
            - 2n \cdot (C^{\mathbf{TC}}_{{DC}_e} + C^{\mathbf{TC}}_{{VC}_e}) \\
            &\quad - \mathbf{C}^{\mathbf{TC}^{\star}}_{Processing}(\bm{\mathcal{P}}_{{PL}}, \mathbf{I}_{{S_e^{PL}}}, B_{{\Omega}^{PL}_e}, F_{PL_e}),
        \end{aligned}
        \end{equation*}
        
        \noindent
        where $\mathbf{C}^{\mathbf{TC}^{\star}}_{Processing}: (\mathbb{R}^{\eta}, \mathbb{R}^{N_A}, \mathbb{R}_{\geq 0}, \mathbb{R}_{\geq 0}) to \mathbb{R}_{\geq 0}$ is a function mapping the design parameters of the processing layer ($\bm{\mathcal{P}}_{PL} \in \mathbb{R}^{\eta} \text{ and } \bm{\mathcal{P}}_{PL} \subset \bm{\mathcal{P}}$); the incentives distributed to attract capital securing the processing layer ($\mathbf{I}_{S^{PL}_e} \in \mathbb{R}^{N_A}_{\geq 0} \text{ and } \mathbf{I}_{S^{PL}_e} \subset \mathbf{I}$); the fees paid for processing the execution logic ($F_{PL_e} \in \mathbb{R}_{\geq 0}$); the budget allocated to compensate the operators of the clearing layer ($B_{\Omega^{PL}_e} \in \mathbb{R}$) for the cost of processing the execution logic; and the capital which is attracted to secure the processing layer is a function of the incentives distributed with a total budget of $B_{S^{PL}_e} \in \mathbb{R}_{\geq 0}$ in epoch $e$.\\
        \\
        With an objective of maximising such cost savings, IntraLayer sees to establish and use a unified clearing layer to optimise costs associated with each transactional path in a cluster.
        \begin{align*}
            &\text{Maximise}_{\bm{\mathcal{P}_{PL}}, \mathbf{I}_{S^{PL}_u}} \quad  \mathbb{E} \left[  \sum^{\mathcal{E}}_{u=1} \sum_{TC_u}\text{CostSavings}_u^{TC} \right] \forall TC_u\\
            &\text{Subject to:} \\
        \end{align*}
        \begin{equation}
        \begin{aligned}
            &\sum_{z=1}^{N_A}  I^z_{S^{PL}_u}  \leq B_{S^{PL}_u}, \\
            & \mathbf{D}_{u + \Delta u} = g_{\mathbf{D}}(\mathbf{D}_{0:u}, \mathbf{\kappa}_{0:u}, \dots) \quad \forall \, \mathbf{D} \in \{ \mathbf{I}_{S^{PL}}\},
        \end{aligned}
        \end{equation}

        \noindent
        where $TC_u$ is the number of transaction clusters operational in $u$-th epoch.
    \end{objective}

 \subsection{Economic Yield}
    The preceding section focused on optimising network's efficiency, akin to reducing the cost of transactional paths between agents. In this section, the focus shifts to enhancing the network's economic yield by: (i) building integrations and attracting agents who will desirably utilise the IntraLayer infrastructure, introducing novel transactional paths and elevating the probability of agents realising their aspired economic objectives; and (ii) increasing the capital efficiency of the transaction pathways supported by the infrastructure.
    
    \begin{objective}[Optimal Agent Population]
        IntraLayer seeks to enhance the utility of it's network by adding new agents, as it's effectiveness hinges upon its ability to efficiently facilitate transactions, which can be improved by incrementally onboarding new agents to the system, such that it maximises the likelihood of the network creating efficient transactional paths. As such, there is a positive correlation between the number of agents ($N_a)$ in the network and the network's value\footnote{Metcalfe’s law \cite{yoo2015moore} suggests that the network's value and the potential connections increase quadratically with the number of nodes, while Zipf's law \cite{adamic2002zipf} implies that the network's value is proportional to $N_a \log N_a$.}. Therefore, the network's utility tends to increase (on average) with the addition of each new agent.\\
        \\
        Specifically, if $\mathcal{V}(N_{a}-1)$ represents the value of network in the presence of $N_{a}-1$ agents, s.t. $\mathcal{V}: \mathbb{N} \rightarrow \mathbb{R}$, then the incremental value of adding the $N_a$-th agent is $\Delta \mathcal{V}(N_a) = \mathcal{V}(N_a) - \mathcal{V}(N_{a}-1)$. Similarly, if $\Delta I_{AA}(N_a) \in \mathbb{R}_{+}$ is the marginal cost of acquiring the $N_a$-th agent, then naturally $\Delta I_{AA}(N_a) = I_{AA}(N_a) - I_{AA} (N_{a}-1)$. At a network level, we can state the value and acquisition costs as follows: 
        \begin{equation}
            \mathcal{V}(N_a) = \mathcal{V}(0) + \sum_{k=1}^{N_a} \Delta \mathcal{V}(k),
        \end{equation}
        \begin{equation}
            I_{AA}(N_a) = I_{AA}(0) + \sum_{k=1}^{N_a} \Delta I_{AA}(k)
        \end{equation}
        
        \noindent
        Since the cost of on-boarding a new agent is not just limited to the incentives provided to the agent to join the IntraLayer network, but also includes expenses associated with integrating the infrastructure with the agent, we use  $\mathbf{\Theta}$ to represent a set of vectors capturing acquisition costs across agents and epochs, such that $\Delta I_{AA} (N_A) \in \mathbf{\Theta}_u$.\\
        \\
        In essence, the IntraLayer seeks to solve the following optimisation problem:
        
        \begin{align*}
            &\text{Maximise}_{\mathbf{\Theta}_u} \quad \mathbb{E} \left[ \sum^{\mathcal{E}}_{u=1} \mathcal{V}_u \right] \\
            &\text{Subject to:} \\
        \end{align*}
        \begin{equation}
        \begin{aligned}
            &\sum^{H_u}_{w=1} \Delta I_{AA}(w) \leq B_{{AA}_u},\\
            &\sum^{H_u}_{h=1} \Delta \mathcal{V}(h) \geq \sum^{H_u}_{h=1} \Delta I_{AA}(h), \\
            & \mathbf{D}_{u + \Delta u} = g_{\mathbf{D}}(\mathbf{D}_{0:u}, \mathbf{\kappa}_{0:u}, \dots) \quad \forall \, \mathbf{D} \in \mathbf{\Theta},
        \end{aligned}
        \end{equation}
        
        \noindent
        where $\mathcal{V}_e \in \mathbb{R}$ is the value of the total network; $B_{{AA}_u} \in \mathbb{R}_{\geq 0}$ represents the budget allocated for agent acquisition in $u$-th epoch; and $H_u \in \mathbb{N}$ denotes the number of agents acquired in $u$-th epoch.
    \end{objective}
    
    \begin{objective}[Path Efficiency]
        IntraLayer seeks to remove siloed value and enhance capital efficiency of all agents involved in the system, as an agent's capital which has been committed to facilitate transactions on a particular path cannot be used to support transactions on other paths, leading to fragmentation.\\
        \\
        We quantify this efficiency using a capital efficiency ($KE^{cq}_e\in \mathbb{R}_{\geq 0}$) measure - representing the capital efficiency of transactional path for agent - $a_{cq}$, which is calculated as follows:
       \begin{equation}
            KE^{cq}_e = \frac{\sum^{N_a-1}_{i=1} E^{cq,i}_e}{K^{cq}_e},
        \end{equation}
        
        \noindent
        where $E^{cq,i}_e$ is the economic output generated by agent $a_{cq}$ by transacting and deploying capital in the transactional path formed with agent $a_{i}$; and $K^{cq}_e$ is the capital deployed in the transactional paths by the agent $a_{cq}$.\\
        \\
        At a network level, IntraLayer seeks to maximise the capital efficiency of transactional paths across all epochs and agents, constrained by its incentive budget and targeted network-owned capital composition.
        \begin{align*}
            &\text{Maximise}_{\bm{\mathcal{P}_{KE}}, \bm{I}_{\mathcal{L}_u}^{\prime}, \bm{\mathcal{K}}_{KE_u}} \quad \mathbb{E} \left[ \frac{1}{\mathcal{E}} \sum^{\mathcal{E}}_{u=1} \sum^{N_a}_{i=1} \mathbf{KE}_u^{i} (\bm{I}^{\prime}_{\mathcal{L}_u}, \bm{\mathcal{P}}_{KE}, \bm{\mathcal{K}}_{{KE_u}}, \bm{F}_{{KE_u}}) \right] \\
            &\text{Subject to:} \\
        \end{align*} 
        
        \begin{equation}
        \begin{aligned}
            &\sum^{N_A}_{o=1}  I_{\mathcal{L}_{{o}_u}}^{\prime} \leq B_{\mathcal{L}_u}^{\prime},\\
            &\sum^{N_A}_{o=1}  \mathcal{K}_{{KE}_{o_u}} = \sum^{N_A}_{o=1}  \mathcal{K}_{{KE}_{o_{u-1}}} + B_{{KE}_{\mathcal{K}_u}},\\
            & \mathbf{D}_{u + 1} = g_{\mathbf{D}}(\mathbf{D}_{0:u}, \mathbf{\kappa}_{0:u}, \dots) \quad \forall \, \mathbf{D} \in \{ \bm{I}_{\mathcal{L}}^{\prime}, \bm{\mathcal{K}}_{KE}\},\\
        \end{aligned}
        \end{equation}
        
        \noindent
        where $\mathbf{KE}^i \in \mathbb{R}_{+}$ is the capital efficiency function which outputs an agent's capital efficiency across all it's transactional paths; $\bm{\mathcal{P}}_{KE} \in \mathbb{R}^{N_A \cdot \mathcal{E}}$ encapsulates the system design parameters that enable higher capital efficiency for the transactional paths; $\mathbf{I}_{\mathcal{L}}^{\prime} \in \mathbb{R}^{N_A \cdot \mathcal{E}}_{\geq 0}$ is the matrix of incentives distributed to attract liquidity used to promote capital efficiency for agents; $\bm{\mathcal{K}}_{KE} \in \mathbb{R}^{N_A \cdot \mathcal{E}}_{\geq 0}$ is the matrix of capital composition controlled by the network for different epochs deployed to facilitate increased capital efficiency for the network's agents; $\bm{\mathcal{K}}_{KE}$ is a submatrix of $\bm{\mathcal{K}}$; $\bm{F}_{KE} \in \mathbb{R}_{\geq 0}^{\mathcal{E}}$ is the vector of fees charged by the infrastructure for enhancing an agent's capital efficiency across different epochs; $I_{\mathcal{L}_{{o}_u}}^{'} \in \mathbb{R}_{\geq 0} $ is the incentives distributed to attract asset $o$; $B_{\mathcal{L}_u}^{\prime} \in \mathbb{R}_{\geq 0}$ is the epoch-specific incentive budget for attracting multi-asset liquidity to enhance capital efficiency for the agents; $\mathcal{K}_{{KE}_{o_u}} \in \mathbb{R}_{\geq 0}$ is the nominal value of asset $o$ controlled by the IntraLayer used to enhance the capital efficiency for the system's agents; $B_{{KE}_{\mathcal{K}_u}} \in \mathbb{R}$ represents the allocated budget to adjust the capital controlled by IntraLayer applied to enhance capital efficiency in the ecosystem.
    \end{objective}

\subsection{Fiscal Sustainability}
    \begin{objective}[Fiscal Sustainability]
        Sustainability of IntraLayer's success requires fiscal prudence to not just be the focal point of algorithmic decisions, but be asserted in spite of changing dynamics. This means that its expenditure (capital outflows) from the budget allocated for the creation and maintenance of transactional paths, should not outweigh associated revenues (capital inflows) in the long run.\\
        \\
        \textbf{IntraLayer's Costs}\\
        \\
        IntraLayer's cost of operations ($\mathbf{B} \in \mathbb{R}^{11 \cdot \mathcal{E}}_{\geq 0}$), i.e. its capital outflow, is essentially the budget it has allocated to various aspects of the network,
        \begin{equation}
            B_u = B_{\mathbf{S}^{DC}_u} + B_{\mathbf{S}^{VT}_u} + B_{\mathbf{S}^{PL}_u} + B_{{\Omega}^{DC}_u} + B_{{\Omega}^{VT}_u} + B_{{\Omega}^{PL}_u} + B_{\mathcal{L}_u} + B_{\mathcal{L}_u}^{'} + B_{{AA}_u} + B_{{VC}_{\mathcal{K}_u}} + B_{{KE}_{\mathcal{K}_u}}
        \end{equation}
        
        \noindent
        where the budget pertaining to technical infrastructure consists of $B_{{\Omega}^{DC}_u} \in \mathbb{R}_{\geq 0}$, $B_{{\Omega}^{VT}_u}  \in \mathbb{R}_{\geq 0}$, $B_{{\Omega}^{PL}_u}  \in \mathbb{R}_{\geq 0}$ comprises $\mathbf{B}_{\Omega}$; economic security includes $B_{\mathbf{S}^{DC}_u}  \in \mathbb{R}_{\geq 0}$, $B_{\mathbf{S}^{VT}_u}  \in \mathbb{R}_{\geq 0}$, $B_{\mathbf{S}^{PL}_u}  \in \mathbb{R}_{\geq 0}$; liquidity provider compensation comprises $B_{\mathcal{L}_u}  \in \mathbb{R}_{\geq 0}$ and $B'_{\mathcal{L}_u}  \in \mathbb{R}_{\geq 0}$; the cost of network-owned liquidity includes $B_{{VC}_{\mathcal{K}_u}}  \in \mathbb{R}_{\geq 0}$ and $B_{{KE}_{\mathcal{K}_u}}  \in \mathbb{R}_{\geq 0}$; and the agent acquisition cost is $B_{{AA}_u}  \in \mathbb{R}_{\geq 0}$.\\
        \\         
        \textbf{IntraLayer's Revenues}\\
        \\
        IntraLayer's revenues are largely derived from the diversity of its transactional paths, and the fees charged for its services. As such, its revenues ($NF \in \mathbb{R}_{\geq 0}$) can be quantified as follows:        
        \begin{equation}
            {NF}_u = NF_{{DC}_u} + NF_{{VT}_u} + NF_{{VC}_u} + NF_{{PL}_u} + NF_{{KE}_u}
        \end{equation}
        
        \clearpage
        \noindent
        \textbf{Cost-adjusted Revenue Maximisation}\\
        \\
        By integrating the previously outlined costs and revenues, IntraLayer aims to achieve financial sustainability through the following decision variables:
        \begin{itemize}
            \item The fee model, $\mathbf{F} \in \mathop{\mathbb{R}}^{m \cdot \mathcal{E}}_{\geq 0}$, which is a matrix representing the fees levied on customers across different services and epochs. 
            \item The budget allocation matrix, $\bm{B} \in \mathbb{R}^{11 \cdot \mathcal{E}}_{\geq 0}$, which is the matrix that influences the network's transactional paths and the economic output derived from these paths.
        \end{itemize}
        
        \noindent
        Any fee structure should strike a balance between incentivising the use of transactional paths (including associated services) and ensuring long-term sustainability, as excessive incentives could jeopardise the network's long-term objectives. Concurrently, when setting the budget allocation matrix, emphasis should be placed on maintaining the reliability of existing transactional paths and encouraging the development of new ones.\\
        \\
        We seek to address the aforementioned problem using the following optimisation statement:
        \begin{align*}
            &\text{Maximise}_{\bm{B}_u, \bm{F}_u} \quad \mathbb{E} \left[ \sum^{\mathcal{E}}_{u=1} \left( \bm{SR}_u(\bm{B}_u, \bm{F}_u) - \alpha \left( \Gamma_u - R_u \right) \right) \right] \\
            &\text{Subject to:} \\
        \end{align*}
        \begin{equation}
        \begin{aligned}
            &R_u \geq \Gamma_u,\\
            & \mathbf{D}_{u + \Delta u} = g_{\mathbf{D}}(\mathbf{D}_{0:u}, \mathbf{\kappa}_{0:u}, \dots) \quad \forall \, \mathbf{D} \in \{ \bm{B}, \bm{F} \},
        \end{aligned}
        \end{equation}
        


        \noindent
        where $\mathbf{SR}_u: (\mathbb{R}^{11 \cdot \mathcal{E}}, \mathbb{R}^{m \cdot \mathcal{E}}) \to \mathbb{R}$, is the system's revenue; $\mathcal{E}$ represents a system parameter defining the duration (in epochs) within which the system seeks to maintain fiscal sustainability; $\mathbf{F}_u$ is the vector of epoch-specific fees charged for different services; and $\alpha \in [0,1]$ is the weighting factor between the objectives of system revenue maximisation and targeted resource management.
    \end{objective}

    \noindent
    Simply put, IntraLayer's business model must incentivise the establishment of new transactional paths and the enhancement of existing paths' quality, through cost reduction and enhanced capital efficiency.
    
 \section{System Architecture}\label{sec:sys}
\begin{figure}[H]
    \begin{center}
        \includegraphics[scale=0.2]{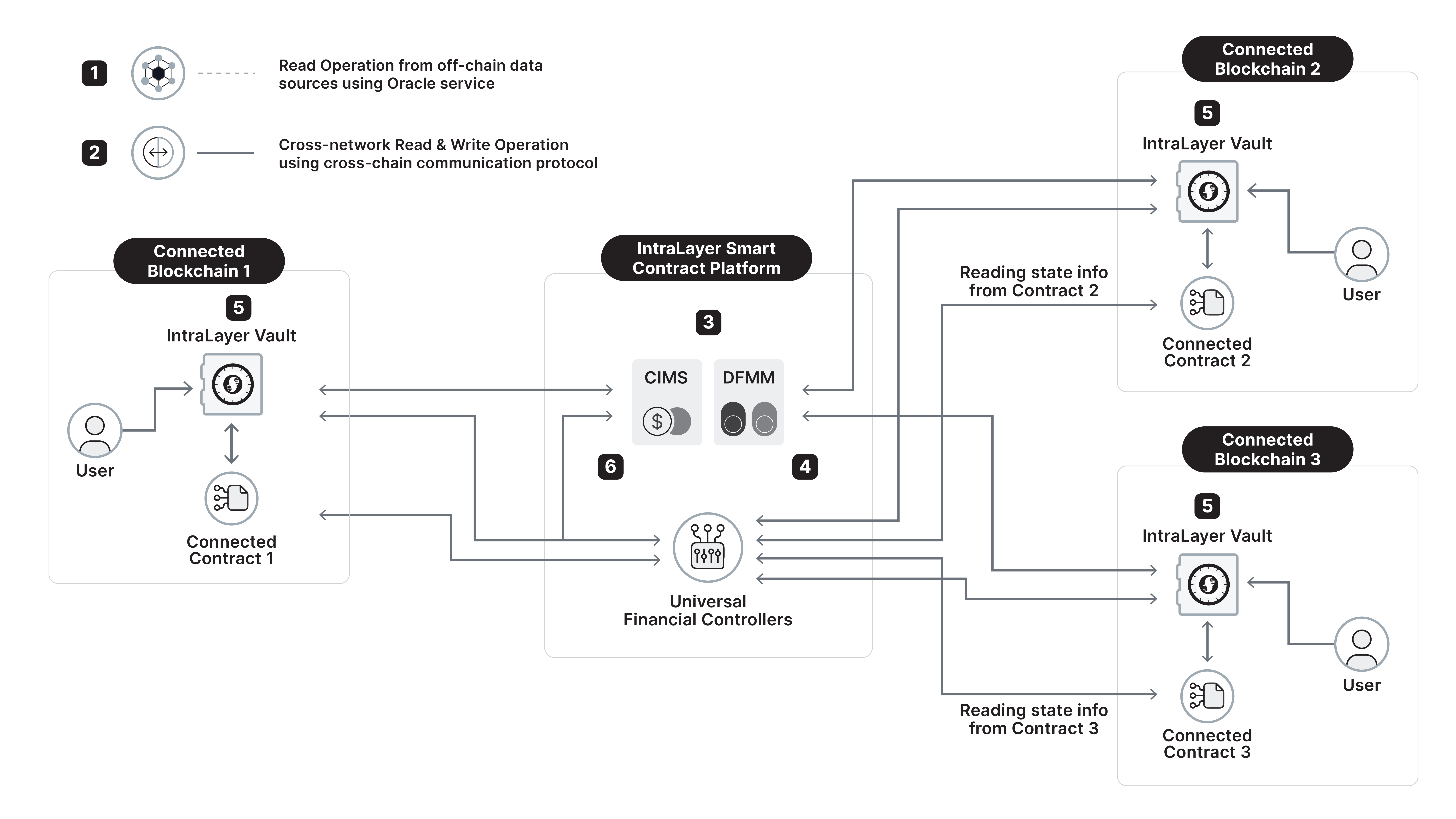}
        \caption{IntraLayer building blocks.}
        \label{fig:schematic}    
    \end{center}
    \end{figure}

    IntraLayer is a multi-purpose, PoS consensus-based decentralised network. It is designed as a key aggregator and router for data, value, and liquidity in the evolving financial landscape, creating transactional paths using a star topology framework, where validator nodes provide vertically integrated services, underpinned by IntraLayer's robust design principles.\\
    \\
    IntraLayer aims to serve as a decentralised gateway, simplifying access to DeFi dApps, users, and other digital finance systems. Its hub-and-spoke design reduces the number of smart contracts needed for cross-platform applications, streamlines asset routing and settlement, and ensures backward compatibility with existing smart contracts. This paradigm allows agents using the IntraLayer network to reach agents on other blockchains and systems, enhancing digital finance interconnectivity. Fig. \ref{fig:schematic} illustrates IntraLayer's architectural components, envisioned to pave the way for the next generation of DeFi systems.
    
    \noindent
    To further elucidate the proposed network's functionality, we describe its infrastructural components, which collectively enable cost-optimised transactional paths between agents.
    
    \begin{enumerate}
        \item \textbf{Oracle Service (OS)}: IntraLayer's oracle service, DORA \cite{chakka2023dora}, establishes secure, reliable, and low-latency connections between blockchains and external systems. Its design, represented by $\bm{\mathcal{P}}_{\text{OS}} \in \mathbb{R}^{\eta} \text{ and } \bm{\mathcal{P}}_{\text{OS}} \subset \bm{\mathcal{P}}_{DC}$, minimises data connectivity costs between off-chain and on-chain systems, achieving Objective 1.
        
        \item \textbf{Cross-chain Communication Protocol (XCP)}: The XCP enables efficient message transmission between connected blockchains, and includes the \textbf{\href{https://supra.com/docs/Supra-HyperNova-Whitepaper.pdf}{HyperNova}} and \textbf{\href{https://supra.com/docs/Supra-Hyperloop-Whitepaper.pdf}{HyperLoop}} protocols, which securely generate, deliver, and verify cross-chain messages, facilitating data and value connectivity. XCP's design, encapsulated by $\bm{\mathcal{P}}_{\text{XCP}} \in \mathbb{R}^{\eta}$, $\bm{\mathcal{P}}_{\text{XCP}} \subset \bm{\mathcal{P}}_{DC}$, and $\bm{\mathcal{P}}_{\text{XCP}} \subset \bm{\mathcal{P}}_{VT}$, reduces data connectivity costs ($C_{{DC}_e}$) and maximises value connectivity efficiency ($CE_{{VT}_e}^{a_{ik}, a_{jk}} \in \mathbb{R}_{+}$), supporting Objectives 1 and 2.
                
        \item \textbf{Smart Contract Platform (SCP)}: The SCP serves as an execution layer, using Moonshot's novel consensus algorithm \cite{doidge2024moonshot}, providing enhanced throughput, reduced block time, and deterministic finality. It enables automated interaction between blockchains and external systems through programmable smart contracts called Universal Financial Controllers (UFCs).
        
        \begin{definition}[Universal Financial Controller]
            A \textit{Universal Financial Controller (UFC)} is a bespoke smart contract deployed on the SCP, containing:
            \begin{itemize}
                \item \textbf{Clearing Logic}: Defines transaction clearing principles and procedures, by formulating and enforcing a comprehensive set of rules governing value, liquidity, and data routing within the ecosystem.
                \item \textbf{State Data Aggregation}: Compiles and analyses state data at set intervals, defined by the UFC developer.
                \item \textbf{Automated Settlement}: Triggers transaction settlements across blockchain networks and off-chain systems.
            \end{itemize}
        \end{definition}

        \noindent        
        UFCs enable IntraLayer's SCP to function as a unified clearing layer for digital finance, facilitating data aggregation and processing from multiple blockchains, ensuring cost-optimised execution of transaction logic ($C^{\mathbf{TC}^{\star}}_{Processing} \in \mathbb{R}_{\geq 0}$), as specified in Objective 6.\\
        \\
        Given IntraLayer's capability to interface with other chains and systems through its oracle service, the XCP, and the IntraLayer vaults, to aggregate and route data and value in a hub-and-spoke fashion across different agents, the network can function as a pivotal gateway entity, reducing complexity and setup cost ($C^{a_n, \mathbf{a}}_{SC} \in \mathbb{R}_{+}$) of enabling interactions among agents across various networks and systems (Objective 4). Working in tandem with UFCs, which enable multi-system and multi-chain transactions, these functionalities collectively ensure streamlined operation and interoperability within the IntraLayer network's ecosystem.
        
        \item \textbf{Dynamic Function Market Maker (DFMM)}: The DFMM is an automated market-making protocol integrated with the liquidity network, acting as an efficient and sustainable price discovery mechanism for single and cross-chain value conversion and transfers by connecting liquidity pools (IntraLayer vaults) across blockchains. By aggregating liquidity and rebalancing open inventory, it facilitates the convergence of an asset's price with its fair market value and reduces execution costs by addressing liquidity fragmentation and minimising slippage. This allows for efficient asset exchange between different agents, reducing liquidity fragmentation and routing complexity through the use of a common counterpart asset, an algorithmic accounting asset, serving as a central node. Overall, the design of the DFMM promotes efficient value exchange.
        
        \item \textbf{Liquidity Network}: IntraLayer's liquidity network consists of interconnected IntraLayer vaults on connected blockchains, forming a network graph $\mathbf{T(V',E')}$, where vertices represent vaults and edges represent the cost of inter-vault asset conversion.
        
        \begin{definition}[IntraLayer Vault]
            An IntraLayer Vault is a smart contract system serving as - (i) liquidity pools to support asset exchange and other financial operations; (ii) digital asset storage, allowing users from various DeFi ecosystems to deposit their assets and access the services facilitated by IntraLayer, functioning as an access point; and (iii) decentralised agent, enabling interaction with and invocation of functions in external smart contracts on the connected blockchain.\\
            \\
            It automates cross-chain transactions by utilising stored gas fee coins to manage gas fee payments efficiently. Its architecture is comprised of three layers, which are as follows:
            
            \begin{itemize}
                \item \textbf{Proxy contract}: Acts as the universal entry point for all executions, ensuring a consistent interface for interactions within the vault.
                \item \textbf{Directory contract}: Maintains a registry of conductors (as described below) and defines parameters for securely validating cross-chain messages.
                \item \textbf{Conductor}: Encapsulates the logic for interaction with external smart contracts.
            \end{itemize}
        \end{definition}
        
        \noindent
        Vaults are integrated with XCP, and through XCP, also with the SCP - which acts as the execution and clearing layer. Through its contract architecture, they are integrated with external smart contracts (Connected Contracts). This integration enables seamless multi-asset value transfers, as IntraLayer can receive data through XCP from external platforms, process data aggregation and execution logic using UFCs, and conduct settlement with the connected contracts through IntraLayer Vaults deployed on the respective blockchains.\\
        \\
        The liquidity network's design, resembling a star graph $\mathbf{T(V^\prime,E^\prime)} \subseteq (\mathbb{N}, \mathbb{N} \cdot \mathbb{N}-1)$, positions the DFMM \cite{abgaryan2023dynamic} (including its intermediating accounting asset), as the central node. This streamlines the hub-and-spoke conversion and exchange of value between nodes, enhancing the flow of liquidity and interoperability of assets across different blockchain networks. The liquidity network acts as a conduit, optimising the cost efficiency of value connectivity between different agents ($CE^{a_{ik},a_{jk}}_{{VT}_e}$), while the DFMM serves as a value conversion engine. The design choices of the DFMM, encapsulated by the parameters ($\bm{\mathcal{P}}_{DFMM} \in \mathbb{R}^{\eta} \text{, and } \bm{\mathcal{P}}_{DFMM} \subset \bm{\mathcal{P}}$), optimise the cost of asset swaps ($CE^{A_{il},A_{jl}}_{{\text{VC}}_e}$), thereby enabling the infrastructure to meet Objectives 2 and 3.
      
        \item \textbf{Cross-chain Inventory Management System (CIMS)}: IntraLayer's CIMS enhances capital efficiency by optimising available collateral utilisation and enabling leveraged transactions across chains, supporting Objective 7.
    \end{enumerate}

\section{Platform of Digital Finance Platforms: An Algorithmic Fiduciary}\label{sec:utility}
    Combining system architecture with business strategy, we envision IntraLayer as a \lq\lq platform of digital finance platforms\rq\rq, serving as an algorithmic fiduciary using decentralised and automated smart contract protocols to faithfully execute agents' (for e.g. dApps, developers, users, etc.) transactional instructions. This can be conceptualised as a meta-system designed to facilitate, execute, and manage interactions across several networks and systems from a single platform.\\
    \\
    IntraLayer is an algorithmic and decentralised system that establishes, maintains, and leverages interconnections within a network of agents, including smart contracts across multiple blockchains and programs in various financial systems. Users codify commands using UFC contracts to initiate conditional transactions with these agents using their assets ($\mathbf{A} = \{A_{11}, A_{12}, \dots, A_{bq}\}$) on different chains, and the platform executes these transactions on behalf of users as an algorithmic fiduciary.\\
    \\
    This intermediation enables agents to deploy and optimally utilise any asset entrusted to the IntraLayer network. It can be envisioned as a multi-chain bank operating under algorithmic reliance. IntraLayer offers users the opportunity to deposit funds in IntraLayer vaults, akin to bank accounts, and optimally use transactional paths facilitated by IntraLayer to deploy their capital and transact across multiple networks.\\
    \\
    For ease of reference, we can liken the role of IntraLayer to that of a bank, by describing its various components:
    
    \begin{itemize} 
        \item \textbf{Head Office} (Smart Contract Platform): Similar to a bank's head office, the IntraLayer smart contract platform establishes overarching governance and security protocols and maintains a master ledger for recording and reconciling transactions across all branches (IntraLayer vaults on other blockchains). It ensures global rules and strategies are enforced, providing centralised oversight akin to a traditional bank's head office.
        \item \textbf{Bank Branches} (Vaults): IntraLayer's vaults on different blockchains function like bank branches, each operating under its unique set of local regulations and customs (similar to different countries' laws). These branches execute instructions from the head office (IntraLayer smart contract platform), ensuring cohesive operation while respecting local blockchain nuances.        
        \item \textbf{Relationship Management} (Smart Contract Interoperability): Much like bank branches forge relationships with local agents, a functionality could be added for IntraLayer Vaults to interface with selected smart contracts on their respective blockchains. These connections allow the network to act on behalf of users, and perform operations and interact with diverse applications, leveraging IntraLayer vaults (bank branches) as operational nodes.
        \item \textbf{Service Agreements} (Universal Financial Controllers): Analogous to customers entering agreements with banks for specific services, UFCs are smart contracts deployed on the IntraLayer network resembling service agreements that users can engage with. They govern multi-chain transactional operations facilitated by IntraLayer, akin to banks conducting transactions with overseas enterprises on behalf of customers.
        \item \textbf{Directives} (Cross-Chain Operations): IntraLayer's network, based on UFCs, sends commands to IntraLayer vaults to manage assets or engage with specific applications, similar to a head office issuing directives to branches. This involves using the cross-chain communication protocol and liquidity network to transfer data and capital based on the \lq\lq service agreement\rq\rq defined in UFCs.
        \item \textbf{Interbranch Capital Transfer} (Liquidity Network \& DFMM): Users may need to move assets across branches (IntraLayer vaults on different blockchains) and applications (connected smart contracts) to engage with other applications on various platforms. This would require a robust price discovery mechanism for efficient asset conversion, similar to currency exchange in a bank - where branches coordinate to use assets in its liquidity network to provide best execution. Similarly, IntraLayer uses its liquidity network and the DFMM protocol to enable inter-branch transactions and ensure fair conversion rates.
        \item \textbf{Audit} (Monitoring): IntraLayer's role in monitoring connected smart contracts and aggregating their state data across blockchains parallels a head office's role in monitoring branch performance, as continuous monitoring would aid in resource allocation and operational optimisation decisions.
        \item \textbf{Developmental Tasks} (Cross-chain Application Development): Just as a bank's head office innovates financial products, the IntraLayer infrastructure evolves through the development of new smart contracts (including UFCs) or products. It provides a platform for multi-chain innovation, enabling developers to create cross-chain applications through a single smart contract deployed on the IntraLayer platform.
    \end{itemize}

    \noindent
    In the schematic that follows, we exhibit how IntraLayer's network uses logic codified in UFCs to govern the bidirectional flow of capital and information across disparate blockchains.
    
    \begin{figure}[H]
    \begin{center}
        \includegraphics[scale=0.2]{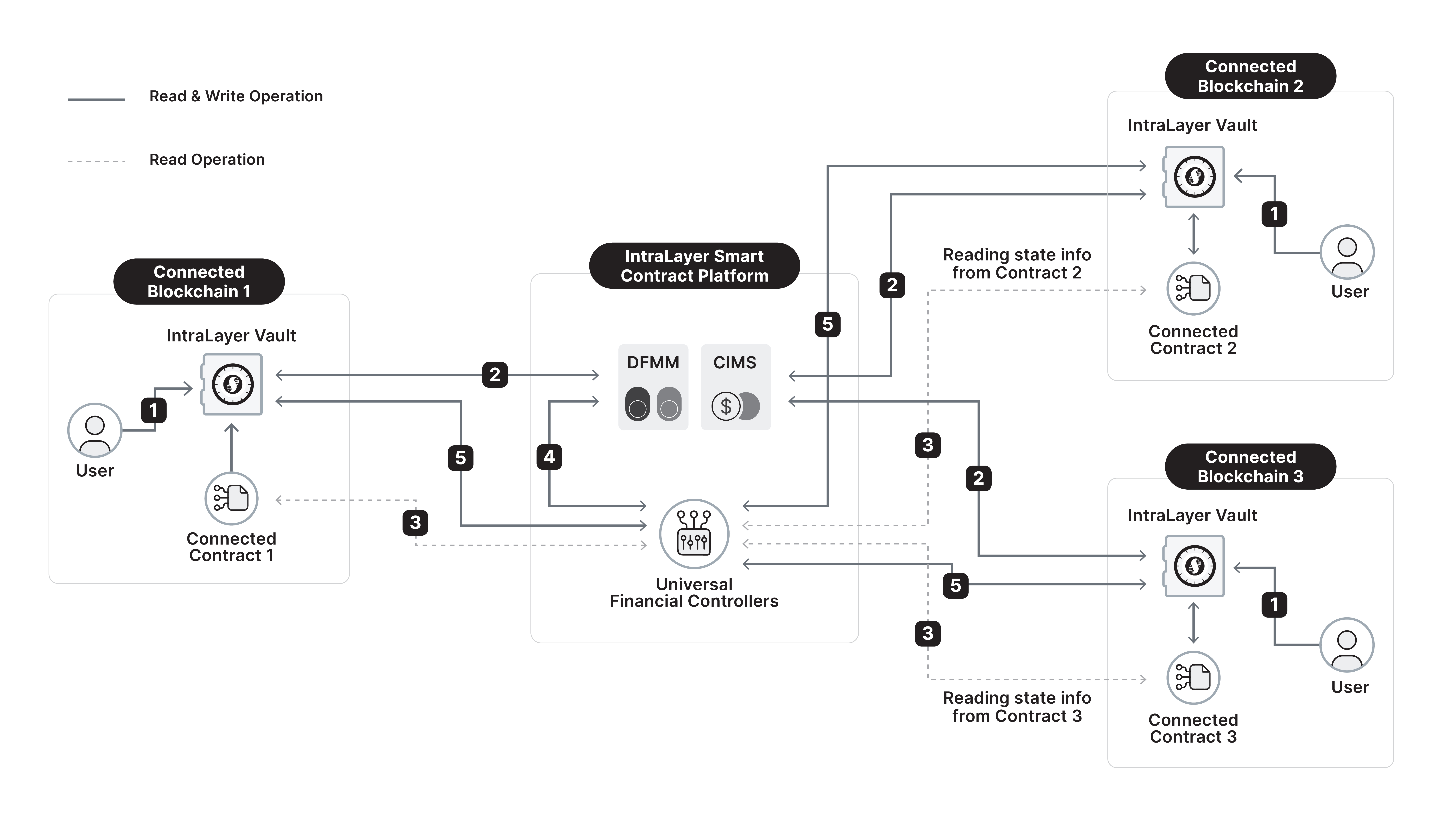}
        \caption{Schematic showcasing operational flow in IntraLayer.}
        \label{fig:IntraLayerdesign}    
    \end{center}
    \end{figure}

    \noindent
     For the reader's reference, we segregate and describe different operations enumerated in the preceding schematic.

    \begin{enumerate}
        \item \textbf{Asset Deposit \& Message Transmission}
        \begin{itemize}
            \item \textbf{Operation}: User deposits assets in the IntraLayer vault on any connected blockchain.
            \item \textbf{Flow}: User posts signed messages (stored on any blockchain or decentralised storage) detailing the specifics (e.g. address, functions) of the UFC they wish to use.
            \item \textbf{Communication}: The IntraLayer cross-chain communication protocol aggregates these messages and vault state data, forwarding them to the designated UFC on the IntraLayer blockchain.
        \end{itemize}
    
        \item \textbf{Interconnected Vault Operations}
        \begin{itemize}
            \item \textbf{Operation}: IntraLayer Vaults, as part of the liquidity network, enable value transition across the liquidity network from the source chain to the target chain.
            \item \textbf{Flow}: Assets are converted into the target network's native assets via the DFMM.
            \item \textbf{Communication}: The conversion and transition of assets are governed by the logic defined in the UFC.
        \end{itemize}
    
        \item \textbf{State Information \& Strategy Determination}
        \begin{itemize}
            \item \textbf{Operation}: The UFC periodically fetches state data from connected contracts through the cross-chain communication protocol, based on pre-defined logic contained within the UFC.
            \item \textbf{Flow}: The UFC can access off-chain data (via Oracle services).
            \item \textbf{Communication}: Based on the ingested data and internal logic, the UFC formulates execution strategies, potentially involving multiple blockchains, and handles accounting and clearing.
        \end{itemize}
    
        \item \textbf{Execution \& Asset Exchange}
        \begin{itemize}
            \item \textbf{Operation}: The UFC's logic may initiate the exchange of assets and orchestrate interactions with connected contracts.
            \item \textbf{Flow}: Utilises the DFMM for price discovery and asset conversion.
            \item \textbf{Communication}: Execution involves IntraLayer Vaults and leverages the liquidity network.
        \end{itemize}
    
        \item \textbf{Routing Decisions \& Assets}:
        \begin{itemize}
            \item \textbf{Operation}: The routing messages are generated based on the execution logic of UFC.
            \item \textbf{Flow}: Routing messages are dispatched to vault contracts using XCP, such that once the message is verified, capital can be promptly transferred to connected smart contracts.
            \item \textbf{Communication}: Vault contracts interact with connected smart contracts based on the validated routing message.
        \end{itemize}
    \end{enumerate}

    \noindent
    IntraLayer's infrastructure supports various operations outlined in the schematic above, ensuring the integrity and consistency of multi-chain interactions. The proposed ecosystem represents a paradigm shift, as automated logic is seamlessly integrated with agents operating across blockchains. This integration not only streamlines capital flow, but also enhances data interoperability, acting as a decentralised \lq\lq message broker\rq\rq. It offers a unified and intuitive interface for users to deploy and interact with cross-chain applications, irrespective of the underlying blockchain architecture.\\
    \\
    Thus far, we have used operations more commonly seen in a bank-like institution to analogise the plethora of use cases utilising IntraLayer's infrastructure. We now focus on how IntraLayer's infrastructure can be instrumental in delivering services similar to the function of a prime brokerage division within a financial institution.
    \subsection{Prime Brokerage}
        We envision IntraLayer's infrastructure to enable efficient transactional settlement across blockchain networks, complemented by a clearing service to manage counterparty risk.\\
        \\
        IntraLayer functions as a decentralised fiduciary entity, orchestrating asset allocation on behalf of users through transactional paths maintained by its infrastructure, thus serving as a transactional conduit within a cross-chain ecosystem.\\        
        \\
        To elucidate the fiduciary responsibility of IntraLayer, consider the scenario illustrated in Figure 3. Initially, users deposit their assets into the IntraLayer vault on Blockchain 3 to engage a specific Universal Financial Controller (UFC). In the second step, following the protocol outlined in the UFC, the system converts the deposited assets into asset $A_{1g}$. Through the IntraLayer vault, which acts as an intermediary executing the smart contract invocation, the funds are transferred to dApp1, deployed on Blockchain 1, for the user's benefit.\\
        \\        
        Subsequently, if a state variable in dApp2 meets certain criteria or upon an additional request by the user, IntraLayer, via the vault contract, is authorised to retrieve assets from dApp1. These assets are then converted into an alternative asset (e.g., $A_{2g}$) through the DFMM and reallocated in dApp2. All these operations are administered by IntraLayer on behalf of the users.\\
        \\
        By entrusting their assets to IntraLayer, users can be assured that subsequent processes with IntraLayer acting as an algorithmic and decentralised fiduciary, solely using the logic codified in UFCs.
        \begin{figure}[H]
            \begin{center}
                \includegraphics[scale=0.25]{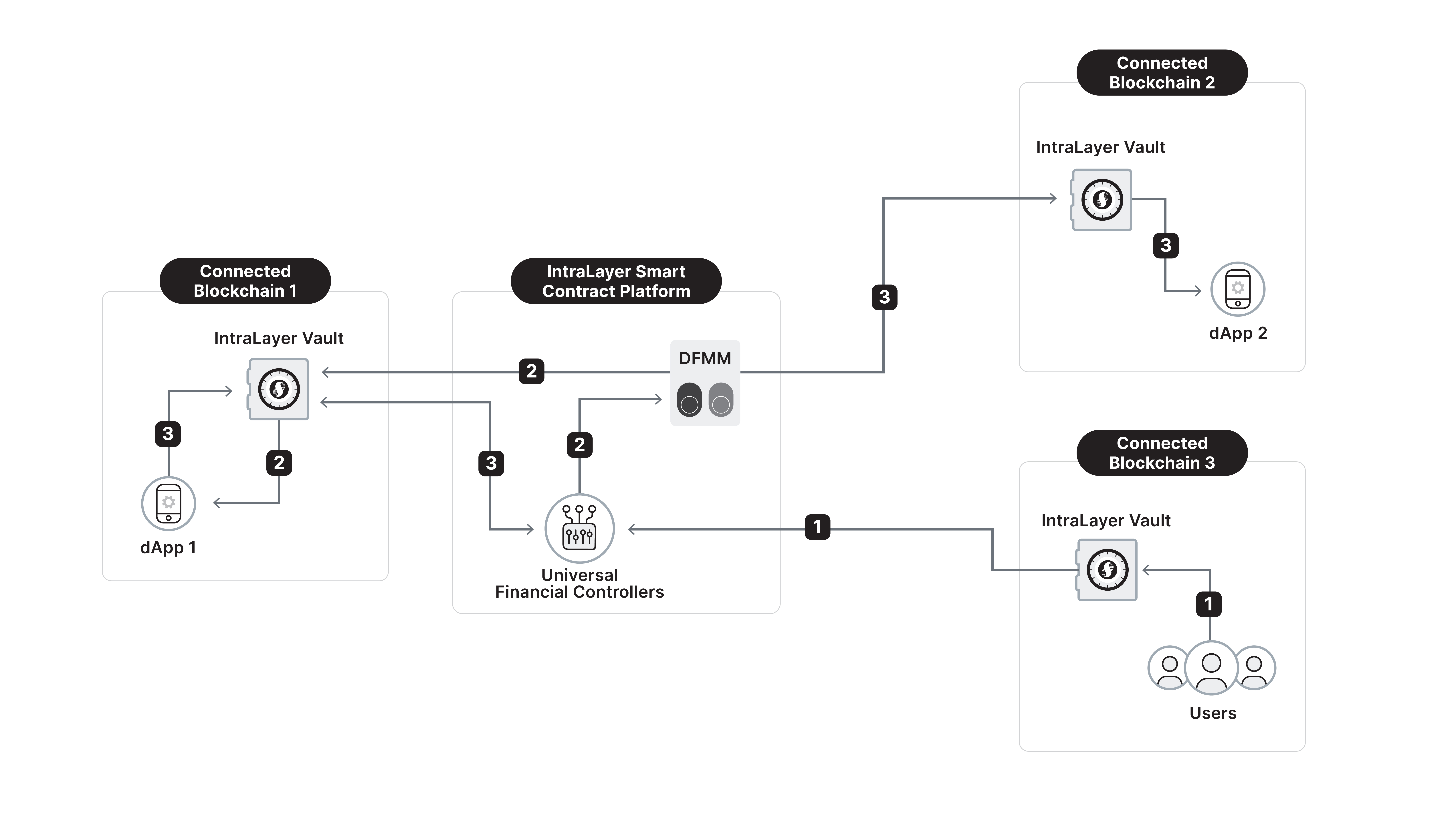}
                \caption{IntraLayer building blocks}
                \label{fig:blocks}    
            \end{center}
        \end{figure}  

        \noindent
        The function of IntraLayer in providing operational support and market access potentially broadens the system's prospects for more advanced services, aligning its functions with those of a traditional prime brokerage service.\\
        \\
        As a prime brokerage service provider, IntraLayer could support the execution of complex financial transactions by utilising established transactional paths. This aims to foster enhanced capital efficiency across transactions facilitated by IntraLayer, aligning with the system's objective of maximising the economic output for its users. This service incorporates a cross-chain inventory management system, which supports economic efficiency in collateral management, using single and cross-chain transactional paths.

    \subsubsection{Cross-chain Inventory Management System}
        IntraLayer's infrastructure seeks to provision a prime brokerage service using its Cross-chain Inventory Management System (CIMS), which seeks to improve the economic efficiency of digital assets deployed with IntraLayer vaults, thereby enhancing the utility of every unit of capital deployed by participants in interconnected blockchains.\\
        \\
        For the reader's convenience, we present CIMS, as part of IntraLayer's infrastructure, using the schematic diagram that follows.
    
        \begin{figure}[H]
            \begin{center}
                \includegraphics[scale=0.25]{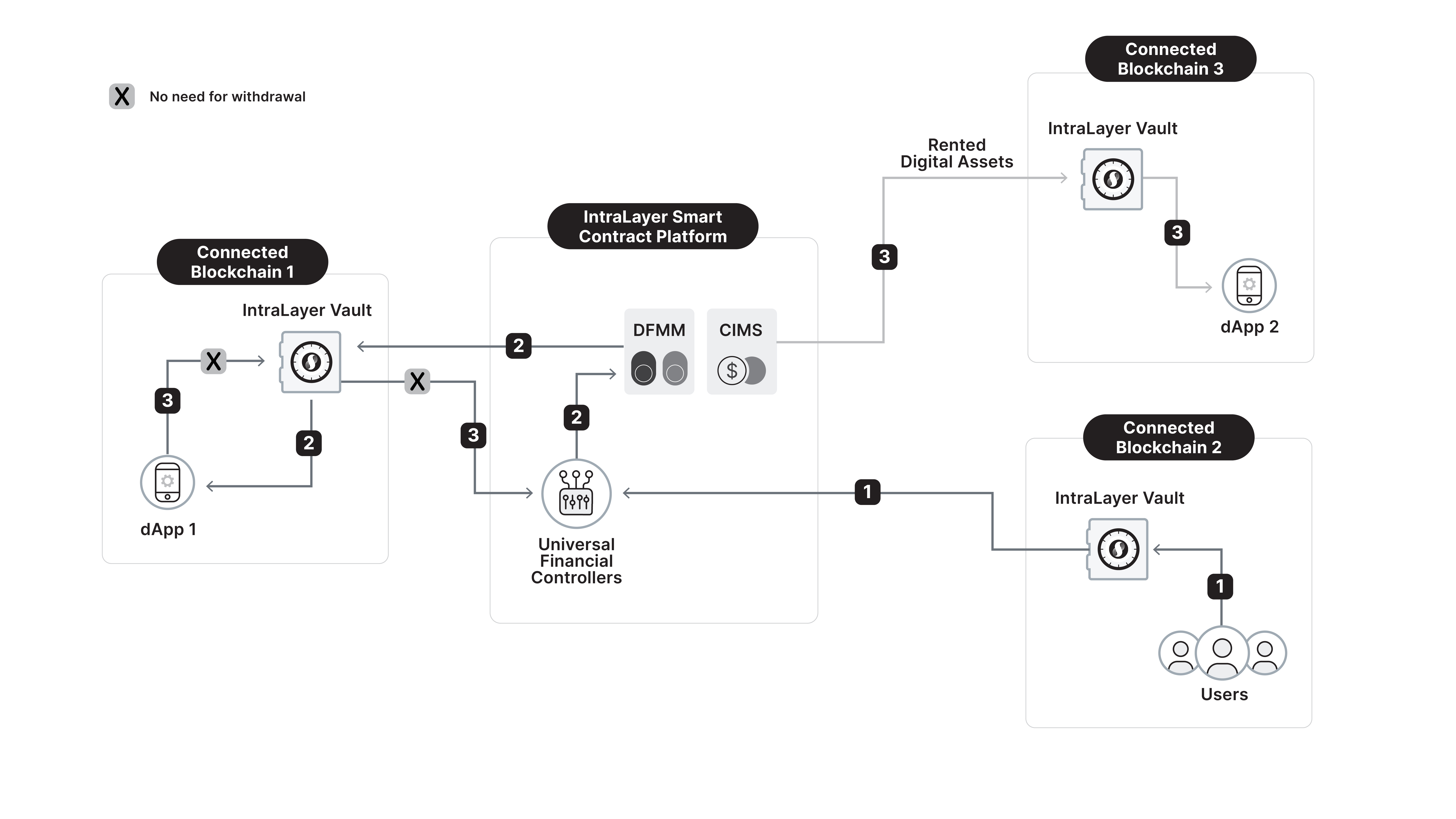}
                \caption{IntraLayer Cross-Chain Inventory Management System.}
                \label{fig:cims}     
            \end{center}
        \end{figure}
    
        \noindent
        Using the logic elucidated in the preceding schematic, once a user allocates capital to dApp1, IntraLayer gains access to state information, signifying that the user has assets in dApp1 controlled by IntraLayer as a fiduciary. These assets can be utilised as collateral to lease assets from the IntraLayer Vaults, enabling users to engage in additional transactional pathways with other agents. For transactions required within dApp2, IntraLayer obviates the need to withdraw assets from dApp1 to facilitate the transaction in dApp2 (as illustrated in Fig. \ref{fig:cims}). Instead, leveraging the assets allocated in dApp1 as collateral, it can lease digital asset inventory to users through CIMS. This allows users to retain their original capital in dApp1 while using leased assets to transact within dApp2, optimising capital efficiency. This design enables IntraLayer to position itself as the network of choice for best execution (as described in Sec. \ref{sec:strategic}), while balancing risks.\\
        \\
        \textbf{Risk Management}\\
        \\
        To mitigate credit and market risks in the event of the lessee's default, IntraLayer ensures the security of the collateral by retaining the right to reclaim leased assets and use the collateral to cover any losses. The system quantifies the required collateral using a collateralisation rate ($\rho_{{A_{1g}}_t}^{a_{1k}, a_{3k}} \in (0,\infty)$), governing the size of inventory a user can access with a fixed amount of asset base in dApp1 ($a_{1k}$), which is calculated as follows:
        
        \begin{equation}
            \rho_{{A_{1g}}_t}^{a_{1k}, a_{3k}} = \frac{H_{{{A_{1g}}}_t}^{a_{1k}} \cdot p_{{A_{1g}}_t} + L^{a_{3k}}_t}{L_t},
        \end{equation}
    
        \noindent
        where $p_{{A_{1g}}_t} \in \mathbb{R}_+$ represents the price of asset $A_{1g}$ at time $t$; $H_{{{A_{1g}}}_t}^{a_{1k}}$ denotes the size of the holdings in asset $A_{1g}$ deployed in the application denoted by $a_{1k}$; $a_{3k}$ denotes the target application where the leased inventory will be deployed; $L_t \in \mathbb{R}_+$ represents the nominal value of the leased digital assets at the point of leasing, and $ L^{a_{3k}}_t \in \mathbb{R}_+ $ denotes the current nominal value of the assets leased that has been submitted to the use-case $a_{3k}$. Whilst at inception $L_t=L^{a_{3k}}_t$, discrepancies will arise if the lessee earns fees (or incurs losses) on the leased assets deployed in application $a_{3k}$.\\ 
        \\
        \textbf{Leveraged Exposure}\\
        \\
        The collateralisation rate can be adjusted to provision, not just over, but also, under collateralised access to inventory ($\frac{{H_{{{A_{1g}}}_t}^{a_{1k}}} \cdot p_{{A_{1g}}_t} + L^{a_{3k}}_t}{L_{t}}<1$), enabling limited leverage for targeted applications. Managed by the CIMS system, the leasing process employs an insurance-based risk assessment framework to mitigate potential losses and determine the optimal collateralisation rate, whether for under or over-collateralised exposure. This framework ensures a seamless user experience, as CIMS operates under the automated logic within UFCs, and by optimising capital efficiency, this mechanism promotes economic output generated from transactional paths, fulfilling objective 7.\\
        \\
        Consider an individual who uses asset $A_{dc}$ as collateral to borrow additional assets in stablecoin $A_{ms}$ through CIMS. The loan can then be used to purchase another asset, $A_{vn}$, with the DFMM, effectively creating leveraged exposure to $A_{vn}$ using the original asset $A_{dc}$, intermediated by $A_{ms}$. This approach serves as an alternative to opening a position in a perpetual futures contract to seek leveraged (long or short) exposures to a specific underlying asset.\\
        \\
        Assuming the nominal value of rented assets is calculated in terms of $A_{ms}$, and considering the user's rented assets at timestep $t$ with collateral size and price remaining the same at timestep $t+1$, the profit and loss (PnL) of leveraged exposure to $A_{vn}$ using $A_{dc}$ as collateral can be calculated as follows:
        
        \begin{equation*}
            \text{PnL}_{t+1} = \left( \frac{C_{{A_{dc}}_t} \cdot p_{{A_{dc}}_t} + L_{t+1}}{\rho^{a_{1k}}_{A_{{dc}_t}}} \right) \cdot \left( \frac{p_{{A_{vn}}_{t+1}}}{p_{{A_{vn}}_t}} - 1 \right)
        \end{equation*}

        \noindent
        where $p_{{A_{dc}}_t}$ is the price of asset $A_{dc}$ at time $t$; $p_{{A_{vn}}_{t+1}}$ and $p_{{A_{vn}}_t}$ are prices of asset $A_{vn}$ at times $t+1$ and $t$, respectively; and $\rho^{a_{1k}}_{A_{{dc}_t}}$ represents the collateralisation rate. The notation, which includes the single agent notation $(a_{1k})$ in the superscript, indicates that after the conversion of rented stablecoin in the DFMM, the converted asset ($A_{vn}$) is deposited into the IntraLayer vault instead of a third-party application. The absence of a superscript in the notation for the leased nominal value implies that the collateral used to rent the asset is also deposited in the IntraLayer vault, rather than in an external application.\\
        \\
        \textbf{dApp \& User Integration}\\
        \\
        Both dApps and individual users can lease assets from the liquidity network, since they can use their capital deployed through IntraLayer as collateral in the CIMS. As an example, AMM protocols deployed using IntraLayer infrastructure can leverage their assets to borrow liquidity, facilitating more competitive services for their users.

\section{The Business Model}\label{sec:business}
    IntraLayer gains its competitive edge by addressing the strategic objectives outlined in Section \ref{sec:obj}, which target a critical gap in the decentralised landscape. We achieve this by deploying UFCs, which act as decentralised programs that coordinate seamlessly with applications across various chains and platforms. This integration enhances multi-chain composability while leveraging IntraLayer's user-centric architecture, positioning it as a key portal for client interactions. As a cross-chain execution layer connected with multiple blockchain systems, IntraLayer significantly expands the reach and versatility of blockchain technology. This unique value proposition not only simplifies user interactions but also solidifies the network's competitive standing in the blockchain industry.\\
    \\
    The remainder of this section covers IntraLayer's phased implementation plan, and outlines various strategic initiatives which we envision would enhance the likelihood of the proposed architecture in meetings its stated objectives.
    \subsection{Phased Implementation}
        The network's lifecycle is delineated into two distinct phases: the bootstrapping phase ($\mathbf{T}_b = [1, 2, \dots, e^\prime]$), and the matured phase ($\mathbf{T}_m = [e'+1, \infty)$). Each phase has specific business objectives and strategies to capture revenue and distribute targetted incentives. As IntraLayer's fundamental utility lies in facilitating and maintaining transactional paths, the business model aims to minimise the cost of transactional paths, sustainably maximise the economic output from these paths, and strategically capture value from provided services as a revenue stream to achieve network objectives.
        \subsubsection{Bootstrapping Phase}
             In the bootstrapping phase, IntraLayer focuses on strengthening its liquidity and security to facilitate efficient transactional dynamics, which are achieved as follows:
            \begin{enumerate}
                \item \textbf{Capital (Liquidity and Staked) Enhancement}: The goal to attract meaningful liquidity and staked capital can positively impact the network, as follows:
                \begin{enumerate}
                    \item \textbf{Cost effectiveness}, which is achieved by enhanced efficiency ($CE_{\text{VC}}$) of value conversion, which is a measure that is calculated as follows:
                        \begin{equation}
                            CE_{\text{VC}} = \sum_{\mu=1}^{e'} \sum_{y=1}^{N_A} \sum_{g=1 \neq y}^{N_A} w_{\mu,y,g} \cdot CE^{A_y, A_g}_{VC_{\mu}},
                        \end{equation}
    
                        \noindent
                        where $CE^{A_y, A_g}_{VC_{\mu}}$ is the efficiency of converting asset $A_y$ to $A_g$ in $\mu$-th epoch; $\mathcal{V}^{A_y, A_g}_{VC_{\mu}}$ is the value converted between the aforementioned assets; $w_{\mu,y,g}$ is the weighting factor for converting value between asset $A_y$ to $A_g$; and the weights $w_{\mu,y,g}$ are calculated as follows:
                        \begin{equation}
                            w_{\mu,y,g} = \frac{\mathcal{V}^{A_y, A_g}_{VC_{\mu}}}{\sum_{y=1}^{n} \sum_{g=1 \neq y}^{n} \mathcal{V}^{A_y, A_g}_{VC_{\mu}}}.        \end{equation}
                    
                    \item \textbf{Security improvement}, which is achieved by enhancing transactional integrity and robustness, using the positive correlation between staked capital and the game-theoretic reliability of IntraLayer network's offerings. Furthermore, improved security also leads to reduction in costs, like: 
                    
                    \begin{itemize}
                        \item Data connectivity cost:
                        \begin{equation*}
                          \sum^{e^{\prime}}_{\mu=1}  \sum_{g=1 \neq l}^{N_a} \sum_{l=1}^{N_a} \left( C^{a_g,a_l}_{\text{DC}_{\mu}} \right),
                        \end{equation*}
                        
                        \item Value transfer efficiency: 
                        \begin{equation*}
                            \frac{\sum^{N_a}_{g=1 \neq l} \sum^{N_a}_{l=1} \left( \mathcal{V}^{g,l}_{VT_e} \cdot CE^{g,l}_{{VT}_e} \right)}{\sum^{N_a}_{g=1 \neq l} \sum^{N_a}_{l=1} (\mathcal{V}^{g,l}_{VT_e})},
                        \end{equation*}
                        
                        \item Processing cost:
                        \begin{equation*}           
                            \sum_{\mu=1}^{e^\prime} \sum_{j=1}^{N_{\mathbf{TC}_{\mu}}} C^{j^\star}_{\text{Processing}}
                        \end{equation*}
                         \noindent
                         where $N_{\mathbf{TC}_\mu}$ is the epoch-specific number of transaction clusters; and $C^{j^\star}_{\text{Processing}} \in \mathbb{R}_{+}$ is the cost of processing transaction cluster $j$,
                    \end{itemize}
                    
                    \item  \textbf{Competitive brokerage}, which is achieved by continuously seeking to enhance the efficiency of providing prime brokerage services, using the cross-chain inventory management system, thereby maximising the capital efficiency ($KE$) of all agents.

                    \begin{equation}
                        KE = \frac{\sum_{\mu=1}^\mathcal{E}  \sum_{g=1}^{N_a} \sum^{N_a-1}_{i=1 \neq g} E^{g,i}_{\mu}}{\sum_{\mu=1}^\mathcal{E} \sum_{g=1}^{N_a} K^{g}_{\mu}},
                    \end{equation}
                \end{enumerate}


            
                \item \textbf{Agent Diversity}: Expanding the network's agent base ($N_a$) exponentially increases the number of potential transactional paths ($N_p \in \mathbb{N}$), necessitating greater diversification of transaction execution and enhancing interconnectivity.
                \begin{equation*}
                    N_p = \kappa \cdot N_a^x
                \end{equation*}
                \noindent
                where $\kappa\in \mathbb{R}^+$ is a proportionality constant, and $x \in (1, \infty)$ reflects the power-law relationship between $N_a$ and $N_p$.
            \end{enumerate}
    
            \noindent
            In the bootstrapping phase, IntraLayer relies on the Proof of Efficient Liquidity (PoEL) protocol \cite{abgaryan2024proof} to attract and retain staked capital and liquidity. The protocol employs a risk structuring engine and incentive allocation strategy to enhance capital efficiency of the on-boarded capital. This protocol introduces non-transferable, time-bound credits, which can be used to offset the cost of using IntraLayer's infrastructural services (such as Oracle services and the smart contract platform). These credits incentivise rapid network engagement by attracting new agents and promoting efficient capital utilisation, including the attraction of new liquidity. This kick-starts a self-reinforcing cycle where higher capital efficiency and more transactional path options attract further capital. During this phase, IntraLayer's primary source of revenue may be fees from DFMM, generated with the amount of assets exchanged, which is enhanced by an increase in the number and quality of transactional paths.
        \subsubsection{Matured Phase}
            Once the IntraLayer network reaches a critical mass of agents and liquidity, signaling its transition to a mature phase, it shifts focus towards consolidating its competitive position and establishing a self-sustaining ecosystem. This involves building robust network-owned liquidity, which can be rented out to agents as part of IntraLayer's prime brokerage services and supplied to DFMM as inventory to more aggressively compete for transactional flow. As the network matures, it will gradually phase out service fee credits and diversify its revenue streams through DFMM, CIMS, and gas fees earned from infrastructural services.
    
\subsection{Strategic Initiatives}\label{sec:strategic}
    Enabling IntraLayer to achieve its true potential would require an integrated strategy that seeks to optimise liquidity management, user acquisition, and revenue distribution, to drive sustainable growth - as enhanced transactional efficiency attracts new liquidity for cross-chain asset conversions and prime brokerage services. We seek to attract multi-asset and productive liquidity by capitalising on IntraLayer's infrastructural novelties, dynamically optimising incentive distribution through our liquidity incentivisation program, and the accumulation of network-owned liquidity.\\
    \\
    IntraLayer is strategically focused on maximising the efficiency of its liquidity utilisation by emphasising cross-chain connectivity, supporting multichain dApp development, and distributing service fee credits to incentivise network usage. This approach fosters increased transaction volumes and maximises the deployment of pooled capital, augmenting user engagement and participation through IntraLayer's infrastructure. The CIMS introduced in Section 5.1.1 enables both individuals and DeFi dApps to lease assets and utilise them within their systems, which incentivises the development of DeFi use cases using IntraLayer by providing better access to capital. Simultaneously, the strategy aims to expand the user base, leveraging network effects to increase usage and liquidity utilisation. This broadened user engagement, coupled with strategic liquidity deployment, catalyses network growth and fosters continuous expansion. Furthermore, the network's fiscal approach supports this growth by efficiently utilising earnings from deployed liquidity and transactional flows, including fees from DFMM, CIMS, and infrastructural use such as XCP, oracle services, and transaction executions. This comprehensive strategy aims to foster a self-sustaining cycle of growth in liquidity and user base, thereby enhancing the network's value and attracting broader interest and engagement.\\
    \\
    Executing this strategy will highlight IntraLayer's distinguishing features and solidify its market position, as meaningful liquidity will attract an expanding user base and transaction volume, securing a dominant market presence by facilitating large transactions with minimal execution costs. Whilst on the other hand, amplified network effects will enhance the network's value, fostering a cycle of ever-growing liquidity and user influx, reinforcing our structural and transactional competitive edge. Over time, by its very design - the network will naturally establish competitive barriers, making it difficult for rivals to replicate the user benefits and infrastructural components provided by IntraLayer, all of which would provide a conducive environment for innovation in cross-platform financial products and services, further enriching the user experience and maintaining IntraLayer's lead in market innovation.
    
    \subsubsection{Towards Network-owned Liquidity}\label{sec:NOL}
        In decentralised exchanges, particularly those using a Constant Function Market Maker model, liquidity providers face substantial risks such as impermanent loss and inventory risk due to positional imbalance. These risks affect not only individual liquidity providers, but also the broader liquidity network. As part of IntraLayer, DFMM addresses these challenges by cultivating a robust liquidity network, strategically deploying capital from passive liquidity providers, and aiming to curtail impermanent loss and inventory risk. This is achieved using an innovative price discovery process, which leverages decentralised oracle services to synthesise price-critical data from diverse platforms into a cohesive virtual order book. This integration diminishes reliance on arbitrageurs to align local market prices with external markets, reducing the impact of market fragmentation. Furthermore, the DFMM framework introduces secondary Liquidity Providers (sLPs) responsible for managing inventory risk, allowing primary Liquidity Providers (pLPs) to delegate this task. This broadens participation and ensures timely rebalancing of inventory levels.\\
        \\
        DFMM provisions the network to productively deploy and sustain network-controlled assets. These assets, defined as network-owned liquidity (NoL), are under the direct ownership and control of IntraLayer without any IOU to external agents or users, and provide a baseline asset portfolio enhancing not only DFMM's, but also the entire network's operational agility and financial management. These assets can be deployed in pLP pools to reduce the need for active or pre-emptive rebalancing, benefiting from DFMM's capacity to mitigate inventory risks arising from unhedged asset exposure, including impermanent loss. This serves the dual purpose of ensuring an enhanced user experience for DFMM through improved cost efficiency in asset conversion and bootstrapping the wider IntraLayer network's services.\\
        \\        
        To gradually build this stable asset base, the system seeks to reserve a portion of network's revenue ($\varpi \in [0,1]$), and channel them as NoL, to be stored in IntraLayer vaults. This asset base's capital efficient (defined using the utilisation rate \cite{abgaryan2023dynamic}) composition is determined by varying demand for various assets, which  is used to find optimal values for of $\bm{\mathcal{K}}_{VC}$, serving as an input to optimisation problems delineated by objectives 3. Additionally, this asset base (optimally comprised of $\bm{\mathcal{K}}_{KE}$) can be used as part of CIMS, further enhancing capital efficiency.
        
    \subsubsection{Proof of Efficient Liquidity}
        The Proof of Efficient Liquidity (PoEL) protocol \cite{abgaryan2024proof} is a helper protocol that is intrinsic to IntraLayer's infrastructure, and supports sustainable liquidity bootstrapping and network security, for specialised Proof-of-Stake (PoS) consensus-based blockchain infrastructures that incorporate intrinsic DeFi applications. This pioneering mechanism judiciously employs staking rewards and fee credits to attract and sustain liquidity from diverse stakeholders, leveraging a risk-conscious incentive allocation strategy to maximise capital efficiency, across multiple epochs.\\
        \\
        In essence, the PoEL protocol seeks to address the optimisation problem outlined below:
        
        \begin{equation*}
        \begin{aligned}
            & \underset{\bm{\zeta},\mathbf{W}}{\text{Maximise}} \; \mathop{\mathbb{E}}[\bm{\xi}_{1:\epsilon}(\bm{\zeta},\mathbf{W})] \\
        \end{aligned}
        \end{equation*}

        \noindent
        where $\epsilon \in \mathbb{R}^+$ denotes the duration of the PoEL's incentivisation program, aligned with the bootstrapping period of the network; $\bm{\zeta} \in \mathbb{R}^{2 \cdot \epsilon}$ represents the set of vectors of total distributed incentives across different epochs (where incentives are in terms of staking rewards and service feecredits); $\mathbf{W} \in \mathbb{R}^{\epsilon \cdot N_A}$ is a matrix of the allocation (\%) of incentives across various liquidity pools (vaults) and epochs; the function $\bm{\xi}_{1: \epsilon}: (\mathbb{R}^{2 \cdot \epsilon}, \mathbb{R}^{\epsilon \cdot N_A}) \rightarrow \mathbb{R}_{+}$ denotes the aggregate capital efficiency of the liquidity network for the bootstrapping period, which maps the reward distribution set of vectors and the incentive allocation matrix to the capital efficiency of the system.\\
        \\
        Whilst the specifics of the PoEL protocol are beyond this paper's scope, it is essential to note that $\bm{\zeta}$ is determined using demand for liquidity across various pools at each epoch, focusing on distributing rewards when capital demand is high and when the use of service fee credits efficiently attracts active use by agents. Similarly, $\mathbf{W}$ is configured to optimally allocate incentives among different pools based on asset demand and the risk characteristics of these assets, as they provide a source of information for the protocol, enhancing its ability to maintain economic security. The overarching aim is to maximise utility for users utilising IntraLayer's transactional paths, fulfilling the network's objectives and facilitating economic value creation.
        
    \subsubsection{iAssets}
        Continuing with IntraLayer's objective to sustainably attract and retain capital by maximising sources, we introduce the concept of an interim Liquidity Provider (iLP).
        
        \begin{definition}[iAsset]
            An iAsset is a specialised asset (issued on IntraLayer's SCP) that LPs receive upon depositing assets into the IntraLayer vault. It represents the holder's ownership in the IntraLayer vault, and enables them to earn rewards, including DFMM and CIMS fees accrued from the utilisation of the deposited assets, as well as staking rewards and service fee credits (enabled with PoEL).
        \end{definition}
        
\begin{figure}[H]
        \begin{center}            
            \includegraphics[scale=0.21]{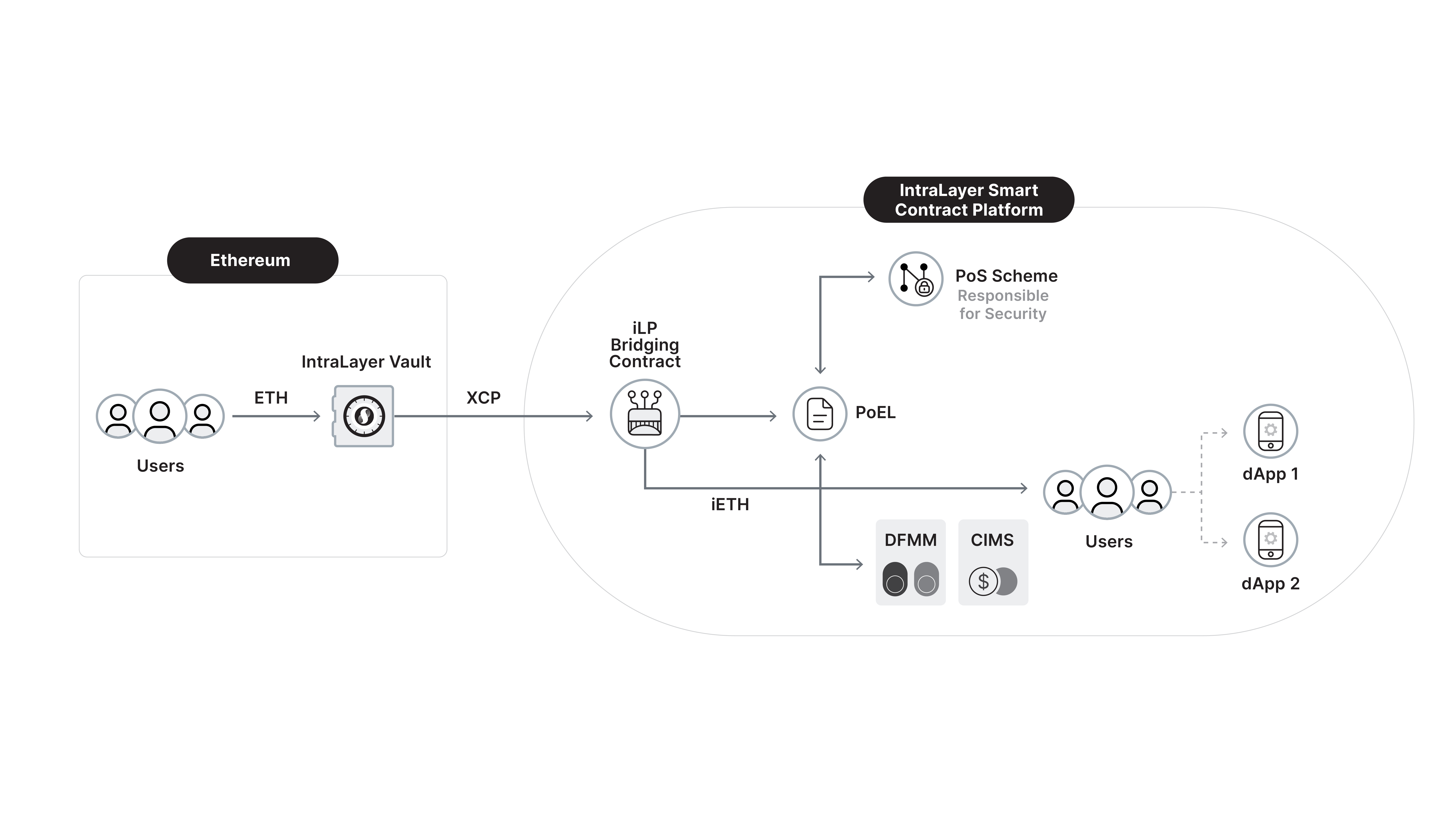}
            \caption{Interm LP Bridging}
            \label{fig:ilp}    
        \end{center}
        \end{figure}

        \begin{definition}[interim Liquidity Provider (iLP)]
            An iLP refers to an LP authorised to manage rights related to their assets on the IntraLayer smart contract platform. This authorisation takes effect once an LP locks their assets (e.g., ETH) in the IntraLayer vault, after which they receive iAssets (e.g., iETH) to manage on the SCP (see Fig. \ref{fig:ilp}).
        \end{definition}

        \noindent
        iAsset holders can manage their assets on the IntraLayer SCP, including using them in DeFi applications and transferring them to other users. Transferring iAssets also conveys the right to receive rewards to the recipient. If an iAsset holder withdraws original assets from the IntraLayer vault or deploys IntraLayer assets for external use, these operations lead to the burning of iAssets. Additionally, to ensure accurate asset accounting in the event of slashing, iAssets could support a rebasing mechanism similar to Ampleforth's implementation \cite{kuo2019ampleforth}, adjusting balances to reflect changes in the original asset's balances in the IntraLayer vault after slashing\footnote{The exact mechanism used to ensure the fungibility of different iAssets by guaranteeing that a slashing event due to validator misbehaviour results in all iAssets of the same type being slashed will be elaborated in the notes that follow.}.\\
        \\ 
        These dynamics are elucidated in the schematic that follows.
   
        \noindent
        Following the acquisition of iAssets (e.g., iETH), users can receive returns from IntraLayer for holding the asset in IntraLayer's native staking asset\footnote{A native staking asset is a intrinsic digital asset of a blockchain network, that is used for locking up a certain amount of capital in order for stakeholders to participate in the network's consensus mechanism.}. If $t_r \in \mathbb{R}_+$ represents the time when an iAsset was received by the user, and $t_s \in \mathbb{R}_+$ represents the time when the user sends the asset to another user, with $t_s > t_r$, the total nominal value of rewards earned by the user $u$ for holding asset $h$ ($\mathcal{R}^{u^h}$) during the period $[t_r, t_s]$, accruing a time-dependent reward rate $r^h_e \in \mathbb{R}_{\geq 0}$, is given by:

        \begin{equation*}
            \mathcal{R}^{u^h} =  \sum^q_{i=1} V^{u^h}_i \cdot r^h_e  
        \end{equation*}

        \noindent
        where $V_{i}^{u^h} \in \mathbb{R}_{\geq 0}$ represents the nominal value of $u$-th user's holdings in iAsset in the vault at $i$-th epoch; $r^h_i \in \mathbb{R}$ represents the rate of return received for the deposited assets $h$, encompassing all factors influencing the rate of return distributed for the asset, such as the staking reward rate, collateralisation rate, and asset-specific demand in the liquidity pools; $q$ is the number of epochs contained within the time interval $[t_r, t_s]$.\\
        \\        
        In summary, this strategic initiative encourages greater LP participation, which helps bootstrap the liquidity network, and promotes utilisation of IntraLayer's SCP, as iAssets can only be managed on the network. Furthermore, if iAssets gain popularity, they will attract more liquidity to IntraLayer pools, making DFMM and CIMS services more competitive, resulting in higher rewards for liquidity providers and increased attractiveness of iAssets. This growth in iAssets will foster an optimal environment for dApps, driving financial activity and the creation of diverse use cases, which will incentivise iAsset holders to retain their assets in the network, thereby perpetuating a cycle of growth using a sustainable source of liquidity for IntrLayer network.

\section{Conclusion and Future Work}\label{sec:conclusions}
    In this work, we have described how IntraLayer's infrastructure can be pivotal in executing a Platform of Digital Finance Platforms vision, using algorithmically reined in smart contracts acting as reliable decentralised fiduciaries. This model is essential for a world that is gradually—and then likely to suddenly—transition to a fully decentralised, efficient, and fully programmatic value exchange system. In the objectives section, we communicated the true essence of our goals using specific optimisation statements, where the decision variables are either system design parameters describing different services of the system, or variables determining the allocation of economic resources of the network. Whilst not solved in this work, these problems were stated with the singular objective of communicating IntraLayer's overarching vision, which encapsulates its protocols and infrastructural components.\\
    \\
    The next iteration of this paper will provide mechanisms and proofs for the optimal allocation of economic resources and fees. We will present a decentralised optimisation framework that facilitates the optimal allocation of the network's economic resources to maximise the probability of IntraLayer achieving its defined objectives. This will be done by modulating the distribution of resources such as service fee credits, incentives for liquidity and staked asset attraction, and other elements of fiscal policy - e.g. service fee, reward rates for service providers, etc. We will test the proposed on-chain optimisation framework by embedding an agent-based simulation into the financial model that encapsulates the system's business model. This will demonstrate that the framework enables the system to achieve its objectives, including long-term sustainability. By simulating several networks, including IntraLayer, and exploring cross-platform transactions handled by IntraLayer, we will show how the system routes value and liquidity between different networks and optimises its operations.

\appendix
\section*{Appendices}
\section{List of Notations}
    \begin{longtable}{|c|p{10cm}|}
        \caption{List of Notations} \label{table:notations} \\
        \hline 
        \textbf{Notation} & \textbf{Description} \\ 
        \hline 
        \endfirsthead      
        \multicolumn{2}{c}
        {{\bfseries Table \thetable\ continued from previous page}} \\
        \hline 
        \textbf{Notation} & \textbf{Description} \\ 
        \hline 
        \endhead        
        \hline 
        \multicolumn{2}{|r|}{{Continued on next page}} \\ 
        \hline
        \endfoot
        \endlastfoot
        
        $\mathbf{G} = (\mathbf{V}, \mathbf{E})$ & Directed graph representing digital finance ecosystem \\ 
        \hline 
        $\mathbf{V}$ & Vertices represent a finite set of agents \\ 
        \hline 
        $\mathbf{E}$ & Edges represent transactional paths \\ 
        \hline 
        $w^{ij}_e$ & Weight representing value or frequency of transactions \\ 
        \hline 
        $O^{a,b}_e$ & Economic output \\ 
        \hline 
        $C^{a,b}_e$ & Function for the cost of a transactional path \\ 
        \hline 
        $f$ & Function mapping costs \\ 
        \hline 
        $\bm{\mathcal{P}}$ & Service-specific design parameters \\ 
        \hline 
        $\bm{I}$ & Tensor of system's incentives (expenditures) \\ 
        \hline 
        $\bm{F}$ & Matrix of fees (revenue) \\ 
        \hline 
        $N_a$ & Number of agents \\ 
        \hline 
        $N_A$ & Number of assets in digital finance ecosystem \\        
        \hline 
        $\mathbf{O^{i,j}_e}$ & Function of economic output for agents $i$ and $j$ at epoch $e$ \\ 
        \hline 
        $R_e$ & Total resources of the system \\ 
        \hline 
        $\Gamma_e$ & Minimum acceptable resources of the system \\ 
        \hline 
        $\mathbf{C^{i,j}_e}$ & Cost Function for agent $i$ and $j$ interaction \\ 
        \hline
        $\bm{\Theta}$ & Vector of sets representing agent acquisition costs \\ 
        \hline 
        $C_{{DC}_e}^{{a_{ik}, a_{jk}}}$ & Cost of data connectivity \\ 
        \hline 
        $\mathcal{D}_{{jk}_t}^{ik}$ & Set of data from agent $a_{ik}$ available to agent $a_{jk}$ \\ 
        \hline          
        $D^{{\star}^{ik}}_{{jk}_t}$ & Optimum set of data from agent $a_{ik}$ available to agent $a_{jk}$ \\ 
        \hline 
        $\Psi^{ik}_{{jk}_t}$ & Discrepancy between received and relevant data \\ 
        \hline 
        $d_{ik, jk}$ & Communication lag \\ 
        \hline
        $\bm{\mathcal{P}}_{{DC}}$ & Vector of parameters for data connectivity \\        
        \hline        
        $\mathbf{I}_{\mathbf{S}^{DC}}$ & Allocation of incentives for data connectivity security \\ 
        \hline 
        $\mathbf{F}_{DC}$ & Fees for data connectivity \\ 
        \hline 
        $\mathbf{B_{{\Omega}^{DC}}}$ & Vector of budget for data connectivity expenses \\ 
        \hline 
        $C_{{VT}_e}^{a_{ik}, a_{jk}}$ & Cost of value transfer \\ 
        \hline 
        $NF_{{VT}_e}^{a_{ik}, a_{jk}}$ & Fee for value transfer \\ 
        \hline 
        $\mathbb{E}[C_{\text{Security}_e}^{a_{ik}, a_{jk}}]$ & Expected security cost of value transfer \\ 
        \hline 
        $\bm{\mathcal{P}}_{VT}$ & Vector of parameters for value transfer \\ 
        \hline 
        $\mathbf{B}_{{\Omega}^{VT}}$ & Vector of budget for value connectivity expenses \\ 
        \hline 
        $\mathcal{V}^{a_{ik}, a_{jk}}_{{VT}_e}$ & Nominal value transferred \\ 
        \hline 
        $\mathcal{V}_{{VT}_e}$ & Total nominal value transferred \\ 
        \hline 
        $CE^{a_{ik}, a_{jk}}_{{VT}_e}$ & Cost efficiency of value transfer \\ 
        \hline 
        $\mathbf{CE}_{{VT}_e}$ & Cost efficiency function for value connectivity \\ 
        \hline 
        $B_{S^{VT}_e}$ & Budget to attract capital to secure value connectivity \\ 
        \hline 
        $\mathbf{F}_{VT}$ & Fees for value transfer \\ 
        \hline 
        $\mathbf{I}_{S^{VT}}$ & Allocation of incentives for data connectivity security \\ 
        \hline        
        $C^{A_{il}, A_{jl}}_{{\text{VC}}_e}$ & Conversion cost \\ 
        \hline 
        $\mathcal{V}^{A_{il}, A_{jl}}_{{\text{VC}}_e}$ & Converted volume \\ 
        \hline 
        $\mathcal{V}_{{VC}_e}$ & Total nominal amount converted \\ 
        \hline 
        $CE^{A_{il}, A_{jl}}_{{\text{VC}}_e}$ & Cost efficiency of asset conversion \\ 
        \hline 
        $\bm{I}_{\mathcal{L}}$ & Incentives to attract liquidity \\ 
        \hline 
        $\bm{\mathcal{P}}_{VC}$ & Vector of parameters for value conversion \\ 
        \hline 
        $B_{{VC}_{\mathcal{K}_e}}$ & Budget to achieve targeted network owned liquidity \\ 
        \hline 
        $\bm{\mathcal{K}}_{VC}$ & Network-controlled portfolio of assets for value conversion \\ 
        \hline 
        $\bm{F}_{VC}$ & Fees for value conversion \\ 
        \hline 
        $B_{\mathcal{L}_e}$ & Incentive budget applied to attract liquidity\\ 
        \hline         
        $\bm{\mathcal{P}}_{SC}$ & System design parameters for setup complexity \\ 
        \hline 
        $\mathbf{CE}_{\text{VC}_e}$ & Cost efficiency function for value conversion \\ 
        \hline 
        $C^{a_{ik}, a_{jk}}_{{SC}}$ & Setup cost \\ 
        \hline        
        $C^{\mathbf{TC}}_{{Clearing}_e}$ & Cost of clearing transaction cluster \\ 
        \hline 
        $C^{\mathbf{TC}}_{{DC}_e}$ & Data communication cost for transaction cluster \\ 
        \hline 
        $C^{\mathbf{TC}}_{{VC}_e}$ & Value transfer cost for transaction cluster \\ 
        \hline         
        $C^{\mathbf{TC}}_{{Processing}_e}$ & Processing cost for transaction cluster \\ 
        \hline        
        $\mathbf{C}^{\mathbf{TC}^{\star}}_{Processing}$ & Function of cost of processing by the unified clearing layer \\ 
        \hline 
        $\bm{\mathcal{P}}_{PL}$ & Design parameters for processing layer \\ 
        \hline 
        $\mathbf{I}_{S^{PL}_e}$ & Incentives for processing layer security \\ 
        \hline 
        $F_{PL_e}$ & Fees for processing layer services \\ 
        \hline 
        $B_{\Omega^{PL}_e}$ & Budget for processing layer expenses \\ 
        \hline 
        $\mathcal{E}$ & System parameter for epochs \\ 
        \hline 
        $B_{S^{DC}_e}$ & Budget for data connectivity security \\ 
        \hline 
        $B_{S^{PL}_e}$ & Budget for processing layer security \\ 
        \hline 
        $B_{\mathcal{L}_e}^{'}$ & Additional budget for compensating liquidity providers \\ 
        \hline 
        $B_{{KE}_{\mathcal{K}_e}}$ & Budget for adjusting NoL, used to enhance capital efficiency \\ 
        \hline 
        $\bm{SR}_e$ & Function of system revenue \\ 
        \hline 
        $\bm{B}$ & Budget allocation matrix \\ 
        \hline 
        $\mathbf{F}$ & Fees levied on customers \\ 
        \hline 
        $R_u$ & Resources of the system \\ 
        \hline 
        $\Gamma_u$ & Targeted resource change \\ 
        \hline 
        $\bm{\mathcal{P}}_{\text{OS}}$ & Vector of design parameters for Oracle Service \\ 
        \hline
        $\bm{\mathcal{P}}_{\text{XCP}}$ & Vector of design parameters for Cross-chain Communication Protocol \\ 
        \hline
        $C_{{DC}_e}$ & Cost of data connectivity \\  
        \hline
        $C^{a_{ik}, a_{jk}}_{{SC}}$ & Setup cost for establishing transactional paths \\ 
        \hline
        $\mathbf{T(V^\prime,E^\prime)}$ & Liquidity network graph with vertices and edges \\ 
        \hline
        $\bm{\mathcal{P}}_{DFMM}$ & Vector of design parameters for Dynamic Function Market Maker \\ 
        \hline
        $\bm{\mathcal{P}}_{KE}$ & Vector of design parameters for capital efficiency \\ 
        \hline
        $\mathbf{KE}^i$ & Capital efficiency function for an agent \\ 
        \hline
        $\bm{\mathcal{P}}_{KE}$ & Matrix of NoL applied to increase capital efficiency \\ 
        \hline
        $\bm{F}_{KE}$ & Vector of Fees charged for enhancing capital efficiency \\ 
        \hline
        $\bm{\mathcal{K}}_{KE}$ & Matrix of allocated capital for capital efficiency \\ 
        \hline
        $CE_{\text{VC}_e}$ & Value conversion efficiency \\ 
        \hline
        $N_p$ & Number of transactional paths \\ 
        \hline
        $\mathbf{T}_b$ & Bootstrapping phase time period \\ 
        \hline
        $\mathbf{T}_m$ & Matured phase time period \\ 
        \hline
        $\kappa$ & Proportionality constant \\ 
        \hline
        $x$ & Power-law relationship exponent \\ 
        \hline
        $\epsilon$ & Duration of the PoEL incentivisation program \\ 
        \hline
        $\bm{\zeta}$ & Set of vectors representing total distributed incentives \\ 
        \hline
        $\mathbf{W}$ & Matrix of incentive allocation across pools and epochs \\ 
        \hline
        $\bm{\xi}_{1:\epsilon}$ & Capital efficiency function for the liquidity network \\ 
        \hline 
        $\varpi$ & Proportion of network revenue reserved for NoL \\ 
        \hline
        $\bm{\mathcal{K}}  $ & Matrix representing NoL asset composition\\ 
        \hline
        $t_r$ & Time when an iAsset was received by the user \\ 
        \hline 
        $t_s$ & Time when the user sends the asset to another user \\ 
        \hline 
        $\mathcal{R}^{u^h}$ & Total nominal value of rewards earned by the user for holding asset $h$ \\ 
        \hline 
        $V_{i}^{u^h}$ & Nominal value of $u$-th user's holdings in iAsset in the vault at $i$-th epoch \\ 
        \hline 
        $r^h_i$ & Rate of return received for the deposited assets $h$ \\ 
        \hline 
        $q$ & Number of epochs contained within the time interval $[t_r, t_s]$ \\ 
        \hline
        $\mathbf{V}$ & Function generating the value of the network based on the number of agents \\ 
        \hline
        $\mathbf{\Theta} $ & Vector of costs associated with onboarding new agents \\ 
        \hline      
        $\mathbf{B}$ & Intralayer's cost of operations \\ 
        \hline
        $\mathbf{\kappa}$ & A set of state variables at a particular epoch \\ 
        \hline
        $g_{\mathbf{D}}$ & A trasnsition function evolving decision variables. \\ 
        \hline
    \end{longtable}

\section{Envisioning a DeFi-CeFi Gateway}
    The IntraLayer infrastructure is uniquely positioned to serve as a pivotal gateway bridging decentralised finance (DeFi) and centralised finance (CeFi), connecting on-chain and off-chain agents and harmonising these distinct financial ecosystems. Its design promotes seamless interoperability and efficient capital flow across diverse financial landscapes, avoiding static pivots that could pose existential risks to the network.\\
    \\
    The IntraLayer architecture addresses several key challenges in data and value transfer, which are encapsulated as follows: 
    \begin{enumerate}
        \item \textbf{Data Exchange}
        \begin{itemize}
            \item \textbf{Gateway Reliability}: Secure and low-latency connections are essential between CeFi and DeFi systems. The Oracle service within the Interlayer provides a reliable gateway for secure and consistent connections between off-chain and on-chain systems.
            \item \textbf{Semantics Interoperability}: A semantic compatibility layer is required to enhance communication efficiency and ensure accurate interpretation of messages between on-chain and off-chain agents. The IntraLayer network is engineered to facilitate semantic interoperability across various blockchains, enabling decentralised creation of compatible messages, such as transforming state updates into API calls (based on automation logic defined in UFCs) or vice versa. This process is authenticated by the network to verify correctness.
            \item \textbf{Policy Enforcement Engine}: Establishing a policy declaration and enforcement layer is crucial for supporting compliant interactions with regulated agents. The smart contract platform within the IntraLayer, through UFCs, allows for the specification of communication rules, which are then trustlessly adhered to by nodes of the network(that facilitate the Oracle service). This framework can extend to any form of policies set by regulators or agents.
     \end{itemize}
     \item \textbf{Value Exchange}
     \begin{itemize}
        \item \textbf{Asset Transfer}: The IntraLayer architecture facilitates the transfer of assets between DeFi protocols and CeFi platforms, ensuring value connectivity across both ecosystems. Non-discretionary fiduciary assistants, off-chain service providers acting on behalf of the IntraLayer network, could execute transactions with off-chain agents on behalf of on-chain agents. These fiduciary assistants act as bridges, facilitating value transfers based on logic and events emitted from UFCs. Their honest behaviour is guaranteed by staking and insurance mechanisms, mitigating risks of misconduct.
        
        \begin{figure}[H]
        \begin{center}
            \includegraphics[scale=0.21]{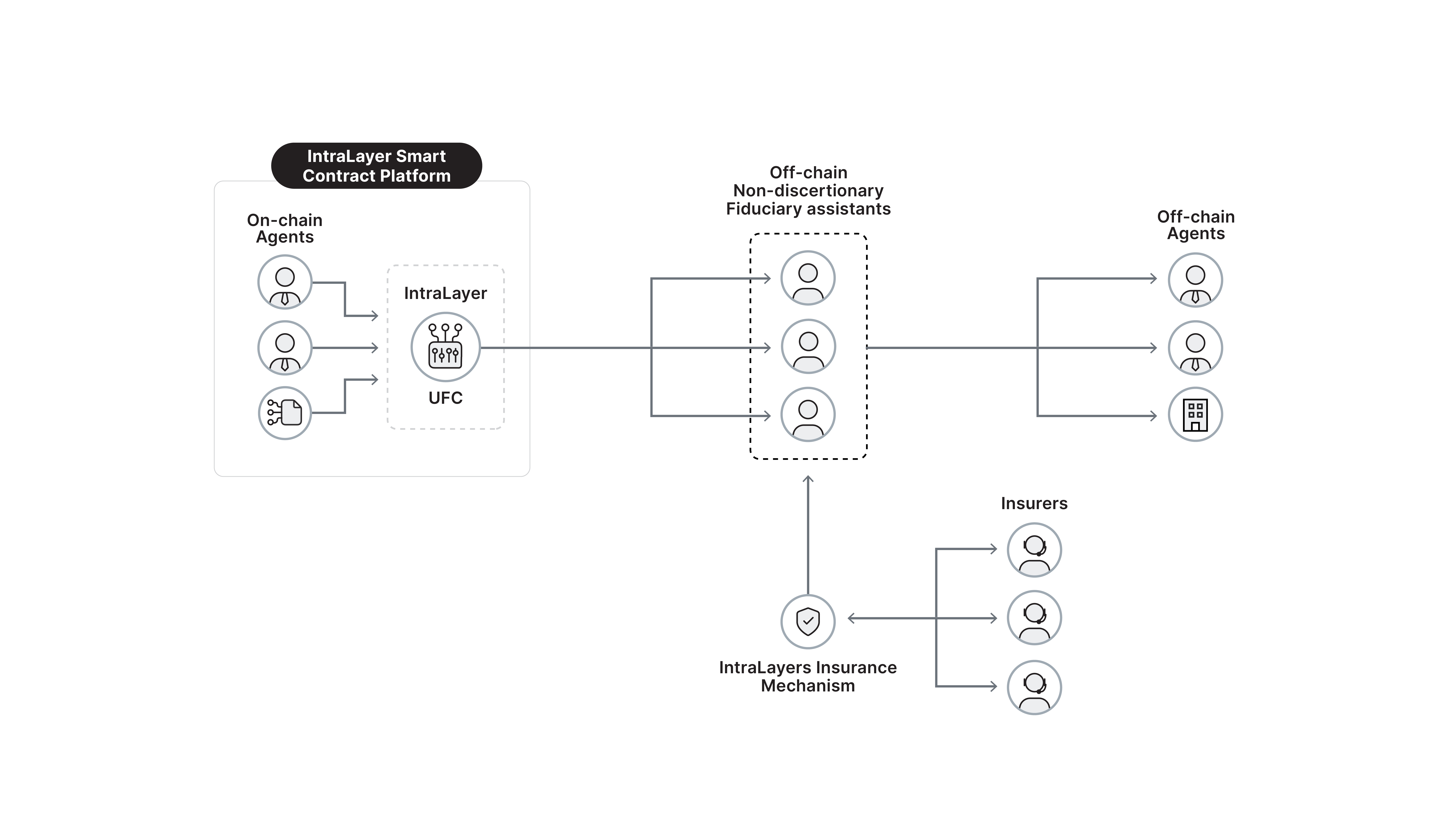}
            \caption{Integrating Off-chain Fiduciary Services.}
            \label{fig:offchainfidu}    
        \end{center}
        \end{figure}
        
        \item \textbf{Risk Management}: In line with the aforementioned insurance mechanism, the IntraLayer architecture is embedded with a robust risk management protocol, which seeks to balance counterparty risk with economic fundamentals. Whilst a full description of this protocol is outside the scope of this paper, we envision this to help us seamlessly connect distinct investment landscapes.
                
        \item \textbf{Market Access}: The infrastructure's optimal interoperability provides DeFi protocols with access to traditional financial markets, expanding investment opportunities and services for users. UFCs can manage both on-chain and off-chain financial transactions, making them truly \lq\lq Universal Financial Controllers\rq\rq.
    \end{itemize}
    \end{enumerate}

    \noindent
    Having broadly considered how data and value exchange can be seamlessly facilitated by the IntraLayer infrastructure, it must also be noted that its use as a gateway does not just enable on-chain to off-chain agent interaction, but also off-chain to off-chain (trustless) interactions. These interactions, which can be facilitated with IntraLayer acting as an integration and verification layer can be broadly categorised as follows:

    \begin{itemize}
        \item \textbf{Decentralised Workflow Logic}: UFCs can embed inter-organisational business process logic, ensuring decentralised execution and verification. This logic is governed in a trustless manner by participating agents, providing a unified plug-and-play environment for record integration and synchronisation among diverse agents.
        \item \textbf{Workflow Integration Bridge}: The IntraLayer acts as a bridge between the UFCs and enterprise applications, linking inter-organisational workflow automation logic to enterprise web services. Oracles facilitate data exchange with legacy systems, such as ERP systems.
        \item \textbf{Reconciliation Layer}: The IntraLayer serves as a reliable clearing, reconciliation, and verification layer, enabling participants to trace contributions and monitor process execution. An integrated identity layer could facilitate verifiable approval of interaction data accuracy between counterparties.
        \item \textbf{Asset Flow Enablement}: Inter-organisational automation processes can be enhanced by the flow of digital assets, both on-chain and off-chain, enabling value exchange between counterparties and DeFi markets through rule-based executions. A gated value migration framework accommodates institutional needs during value exchange.
        \item \textbf{Privacy-Preserving Consistency}: The system enforces verifiable consistency between parties' records when necessary, without moving data or business logic from legacy systems, by incorporating privacy-preserving technologies such as ZK proofs \cite{almashaqbeh2012}.
    \end{itemize}

    \noindent
    Additionally, the IntraLayer infrastructure can facilitate regulatory compliance through a rules-based compliance layer, addressing another challenge faced by CeFi institutions when assessing DeFi infrastructure for executing traditionally centralised operations. Thus, the IntraLayer infrastructure not only bridges DeFi and CeFi, but also enhances capital efficiency for both ecosystems, providing access to larger liquidity pools, broadening market access, and facilitating seamless price discovery for a more integrated and efficient financial landscape.\\
    \\
    For ease of reference, the schematic which follows seeks to illustrate how an off-chain entity, such as a bank, can leverage the IntraLayer system as a gateway for integration and automation of interactions with blockchain networks and other off-chain systems. The bank deploys UFCs within the IntraLayer to facilitate interactions with various agents. For illustration purposes, a gatekeeper (either the bank itself or a designated service provider) could manage identity and whitelist third-party developers. These developers would be authorised to create additional smart contracts that interact with the bank's system, promoting open innovation. The bridge and oracle service could enable data and value routing by the bank to other on-chain and off-chain agents, using the bank's UFCs to define automation and clearing logic.
    
    \begin{figure}[H]
    \begin{center}
        \includegraphics[scale=0.21]{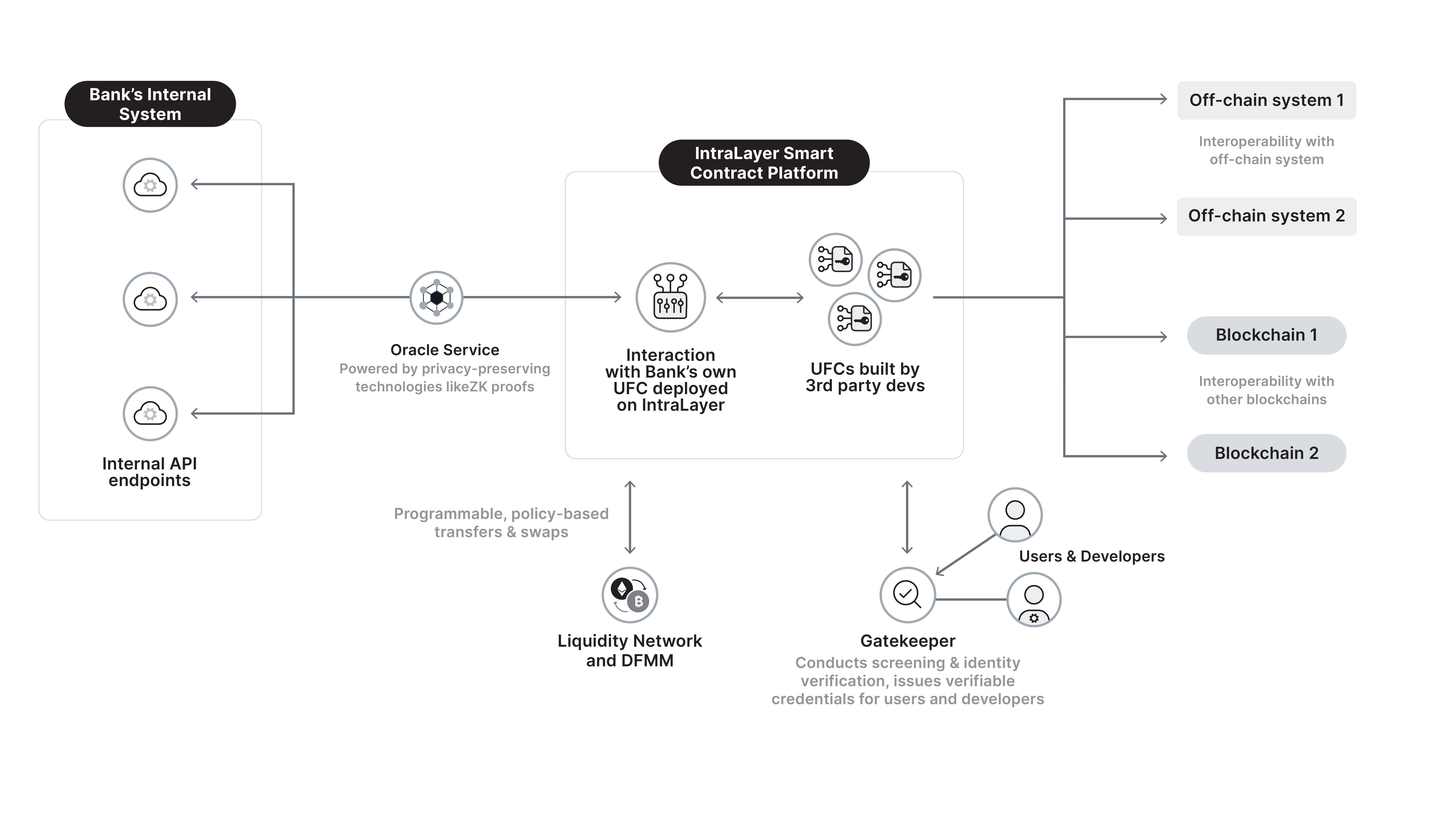}
        \caption{IntraLayer Example: Secure an ecosystem of services interoperable with its legacy infrastructure.}
        \label{fig:intralayer_design}    
    \end{center}
    \end{figure}

    \bibliography{main.bib}
    \bibliographystyle{plain}
    
\end{document}